\documentclass[twocolumn,apj]{emulateapj}
\usepackage{apjfonts}

\newcommand{\beq}{\begin{equation}}
\newcommand{\eeq}{\end{equation}}
\newcommand{\beqn}{\begin{eqnarray}}
\newcommand{\eeqn}{\end{eqnarray}}
\newcommand{\AU}{{\rm AU}}

\newcommand{\kB}{k_{\rm B}}
\newcommand{\si}{{\rm i}}
\newcommand{\se}{{\rm e}}
\newcommand{\sd}{{\rm d}}
\newcommand{\sg}{{\rm g}}

\newcommand{\bracket}[1]{\langle #1 \rangle}
\newcommand{\eqref}[1]{(\ref{#1})}
\newcommand{\dsum}{\displaystyle\sum}
\newcommand{\dint}{\displaystyle\int}
\newcommand{\dfrac}[2]{ {\displaystyle\frac{#1}{#2}} }

\newcommand{\pfrac}[2]{ \Bigl(\dfrac{#1}{#2}\Bigr) }

\shorttitle{ELECTRIC CHARGING OF DUST IN PROTOPLANETARY DISKS}
\shortauthors{OKUZUMI}
\slugcomment{Sumitted to ApJ 2009 January 20; accepted 2009 April 13}

\begin{document}
\title{Electric charging of dust aggregates 
and its effect on dust coagulation in protoplanetary disks}
\author{Satoshi Okuzumi}
\affil{Graduate School of Human and Environmental Studies, Kyoto University,
Yoshida-nihonmatsu-cho, Sakyo-ku, Kyoto 606-8501, Japan}
\email{satoshi.okuzumi@ax2.ecs.kyoto-u.ac.jp}
\begin{abstract}
Mutual sticking of dust aggregates is the first step toward planetesimal 
formation in protoplanetary disks.
In spite that the electric charging of dust particles is well recognized in some contexts,
it has been largely ignored in the current modeling of dust coagulation.
In this study, we present a general analysis of the dust charge state in protoplanetary disks,
and then demonstrate how the electric charging could dramatically change the currently accepted
scenario of dust coagulation.
First, we describe a new semianalytical method 
to calculate the dust charge state and gas ionization state self-consistently.
This method is far more efficient than previous numerical methods, and provides a 
general and clear description of the charge state of gas-dust mixture. 
Second, we apply this analysis to compute the collisional cross section of growing aggregates
taking their charging into account.
As an illustrative example, we focus on early evolutionary stages where the dust has been thought to grow into fractal ($D \sim 2$) aggregates with a quasi-monodisperse (i.e., narrow) size distribution.
We find that, for a wide range of model parameters, the fractal growth is strongly inhibited
 by the electric repulsion between colliding aggregates and eventually ``freezes out'' on its way to the subsequent growth stage involving collisional compression.
Strong disk turbulence would help the aggregates to overcome this growth barrier,
but then it would cause catastrophic collisional fragmentation in later growth stages.
These facts suggest that the combination of electric repulsion and collisional fragmentation 
would impose a serious limitation on dust growth in protoplanetary disks.
We propose a possible scenario of dust evolution after the freeze-out.
Finally, we point out that the fractal growth of dust aggregates tends to 
maintain a low ionization degree and, as a result, a large magnetorotationally stable region
in the disk.
% \received{hreceipt datei}
% \revised{hrevision datei}
% \accepted{hacceptance datei}

\end{abstract}
\maketitle

%%%%%%%%%%%%%%%%%%%%%%
\section{Introduction}
%%%%%%%%%%%%%%%%%%%%%%
The initial step toward planetesimal formation in protoplanetary disks 
is the collisional growth of submicron dust grains into macroscopic aggregates.
A standard scenario is that dust aggregates grow by mutual sticking,
gradually settle to the midplane of the disk, and finally form a dense dust layer.
It is still an open issue whether subsequent growth is established 
by the gravitational instability of the layer or the direct growth of the aggregates. 
To address this issue, further understanding on earlier evolutionary stages is needed.

It has been recognized that the internal structure of aggregates is a key factor 
for their growth and settling.
Early studies on dust coagulation modeled the aggregates 
 as a compact, nonporous object (e.g., \citealt*{Weidenschilling80,NNH81}).
Both numerical simulations and laboratory experiments have revealed,
however, that aggregates are not at all compact, but has an open, 
fluffy structure (for a review, see \citealt*{Meakin91,Blum04,DBCW07}).
This is particularly true for aggregates formed at an early growth stage where 
the collisional velocity is too low for colliding aggregates to compress each other.
It has been observed in numerical \citep{Kempf+99} as well as experimental
 \citep{WB98,Blum+98,Blum+00} studies that the outcome is an ensemble of
fractal aggregates with the fractal dimension $D \la 2$ and with a quasi-monodisperse (i.e., narrow) mass distribution. 
This fractal growth typically lasts until the aggregates become centimeter-sized \citep{SWT08}.
A remarkable dynamical property of these fluffy aggregates is that they 
keep a strong coupling to ambient gas and thus a low drift velocity relative to the gas
 throughout the evolution.
This could be crucial to the formation of very thin dust layer where planetesimals 
may be formed by gravitational instability.

Dust grains and aggregates are not only the building block of planetesimals but also 
 powerful absorbers of charged particles in the gas disks.
It is now widely accepted that turbulence in the disks is attributed to 
magnetorotational instability (MRI; \citealt*{BW91}).
For this mechanism to work,
at least a part of the disk needs to be sufficiently ionized for the gas 
to couple to  magnetic fields.
Many authors have examined whether protoplanetary disks 
can be ionized enough to sustain MHD turbulence \citep{Gammie96,GNI97,Sano+00,IG99,IN06,IN06b,IN06c,Wardle07}.
One of the important findings is that the turbulent region is strongly controlled 
by the concentration of dust materials since they efficiently remove away ionized particles 
from ambient gas \citep{Sano+00,IN06,Wardle07}.

Although the importance of dust charging is well recognized in the above context,
its effect on dust coagulation in protoplanetary disks has been hardly examined.
Charging of aggregates causes electrostatic interaction between them, 
which may significantly increase or decrease the coagulation rates.
Recently, a series of studies have suggested that charge-induced dipole interaction 
might trigger runaway growth of dust aggregates \citep{IMK02,Konopka+05}.
However, these studies considered a situation where the ambient gas is not ionized 
and the net charge of dust aggregates vanishes identically.
In protoplanetary disks, on the contrary, the net charge of dust aggregates does {\it not} vanish due to the presence of weakly ionized ambient gas, and therefore both dust charging {\it and} gas ionization
must be taken into account. 

This study explores how the electrostatic charging of dust aggregates 
could be crucial to their coagulation in protoplanetary disks.
For this purpose, we have to know in advance how the charge state of aggregates
 evolves with their growth.
This is a complicated problem, since we also have to solve the
 ionization state of ambient gases self-consistently.
Previous studies \citep{Sano+00,IN06,Wardle07} have handled this problem 
with direct numerical calculations in which dust particles with 
different charges and sizes are treated as different charged species as well as 
many species of ions.
However, this approach becomes inefficient when one tries to solve this problem and dust growth simultaneously,
since the dispersion of charge and size increases as the dust growth. 
The central strategy taken in this study is to solve this problem {\it as analytically as possible}.
This approach does not only reduce the computational expense but also provides 
general insight into the charge state of gas-dust mixture.
As a result, we show that all the conditions for ionization-recombination equilibrium 
are reduced to a {\it single} algebraic equation.
Just by solving this equation numerically, we can obtain both of the dust charge state and
the gas ionization state analytically.
We also confirm that the semianalytical calculations
agree very well with direct numerical calculations using the original equations. 
This semianalytical method will be a powerful tool for the simulations
of charged dust coagulation and MRI turbulence. 

As an illustrative example, we calculate the collisional cross section of dust aggregates
growing in a protoplanetary disk taking into account their electric charging.
We focus on early stages of dust evolution where the aggregates has been thought to experience
fractal, quasi-monodisperse growth (e.g., \citealt*{Blum04,DBCW07}).
For a wide range of model parameters, 
we find that the effective cross section is quickly suppressed as the fractal growth proceeds 
and finally vanishes at a surprisingly early stage.
This means that the fractal growth  ``freezes out'' on its way to the subsequent growth stage where collisional compression of aggregates occurs.
This is because the electrostatic repulsion between aggregates becomes strong enough 
to prevent their mutual collision.
Strong turbulence in the disk will help the aggregates to overcome 
this electric barrier, but then it will cause catastrophic disruption 
of collided aggregates at later stages.
Therefore, if the freeze-out of the fractal growth truly means the end of dust evolution,
the combination of the electric charging and the collisional disruption imposes 
a very strict limitation on dust coagulation and subsequent planetesimal formation 
in protoplanetary disks.
Our findings strongly suggest that the dust charing effect should be seriously 
taking into account in the modeling of dust evolution.

This paper is organized as follows. In \S2, we present a set of equations that
describes the reactions of charged particles (ions, electrons, and dust aggregates),
and derive the equation that determines the equilibrium state.
In \S3, we calculate the electrostatic repulsion energy between two colliding aggregates
to show that the quasi-monodisperse fractal growth is strongly inhibited
for a wide range of disk parameters.
In \S4, we discuss the validity of some important assumptions and point out a possible scenario
of dust evolution after the ``freeze-out'' of the fractal growth. 
A summary is presented in \S5.

%%%%%%%%%%%%%%%%%%%%%%%%%
\section{Equilibrium charge distribution}
%%%%%%%%%%%%%%%%%%%%%%%%%
%%%%%%%%%%%%%%%%%%%%%%%%%%%%%%%%%%%%%%%%%%%%%%%
\subsection{Kinetic equations for ionization-recombination reactions}
\begin{figure}
\plotone{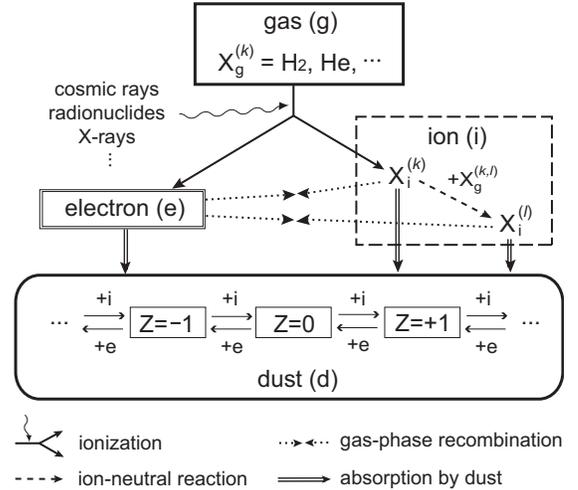}
\caption{Schematic diagram of the ionization-recombination reactions in a gas-dust mixture.
Ions and free electrons are created ({\it solid arrows}) by some ionizing sources (e.g., cosmic rays)
and are removed through the gas-phase recombination ({\it dotted arrows}) 
or the adsorption to dust ({\it double line arrows}). Some species of ions may react with
neutral gas particles to create different species of ions ({\it dashed arrows}).
The equilibrium charge distribution $n_\sd(Z)$ of dust aggregates are determined
by the balance between all these reactions.}
\label{fig:reaction}
\end{figure}

%%%%%%%%%%%%%%%%%%%%%%%%%%%%%%%%%%%%%%%%%%%%%%%
We model the ionization-recombination reactions in a gas-dust mixture as follows
(see also fig.\ref{fig:reaction}).
Some ionizing sources (e.g. cosmic rays) create ions ${\rm X}_\si^{(k)}$ and free electrons
from neutral particles ${\rm X}_\sg^{(k)}$ at a rate $\zeta^{(k)}$ 
(here $k(=1,2,\cdots)$ labels each species of ions and associated neutrals).
We assume that the ions and electrons are quickly thermalized and have thermal velocities
$u_\si^{(k)}$ and $u_\se$, respectively.
We neglect the possibility that the free electrons might be much more energetic 
in a MRI-active region \citep{IS05}.
The ions may react with neutrals ${\rm X}_\sg^{(k,l)}$ to produce another species of ions
${\rm X}_\si^{(l)} \,(l\not=k)$, or may recombine with free electrons in the gas phase.
We denote the rate coefficient for the ion-neutral reaction and the gas-phase recombination 
by $\beta^{'(k,l)}$ and $\beta^{(k)}$, respectively. 
Also, the ions and free electrons may collide with dust aggregates to adsorb onto their surfaces.
We write the collisional cross section for an aggregate and an ion (electron) as $\sigma_{\sd\si(\sd\se)}$.
These cross sections generally include the effect of electrostatic interaction as well as
the sticking probability  (see \S2.2).
Each dust aggregate may have different internal structure and charge $Ze$ from the other. 
We represent a set of parameters describing the structure (e.g., mass, radius) 
as $I = \{I_1,I_2,\cdots\}$.
In this section, we assume that the above reactions proceeds faster than 
the mutual collision of dust aggregates and treat $I$ as constant parameters.
The validity of this assumption is discussed in \S4.1.
%%%%%%%%%%%%%%
%%%%%%%%%%%

The above charge reactions are described by a set of kinetic equations. 
Let us denote the number densities of ions ${\rm X}_\si^{(k)}$, free electrons, 
and dust aggregates as $n^{(k)}_\si$, $n_\se$, and $n_\sd(I,Z)$, respectively.
The rate equations for $n_\si^{(k)}$, $n_\se$, and $n_\sd(I,Z)$ are given by 
\beqn
\dot{n}_\si^{(k)} &=& \zeta^{(k)} n_\sg^{(k)} 
- u_\si^{(k)} n_\si^{(k)} \dint dI \dsum_Z \sigma_{\sd\si}(I,Z)n_\sd(I,Z) 
\nonumber \\
&&- \beta^{(k)} n_\si^{(k)}n_\se 
-\sum_l \left[\beta^{'(k,l)}n_\si^{(k)}n_\sg^{(k,l)}
-\beta^{'(l,k)}n_\si^{(l)}n_\sg^{(l,k)}\right],
\label{eq:evol_nik}
\eeqn
\beqn
\dot{n}_\se &=& \sum_k \zeta^{(k)} n_\sg^{(k)}
- u_\se n_\se \dint dI \sum_Z\sigma_{\sd\se}(I,Z)n_\sd(I,Z)
\nonumber \\
&&- \sum_k \beta^{(k)} n_\si^{(k)}n_\se,
\label{eq:evol_nek}
\eeqn
and 
\beqn
\dot{n}_\sd(I,Z) &=& \sum_k u_\si^{(k)} n_\si^{(k)} [\sigma_{\sd\si}(I,Z-1)n_\sd(I,Z-1)
-\sigma_{\sd\si}(I,Z)n_\sd(I,Z)]    \nonumber \\
&& +u_\se n_\se [\sigma_{\sd\se}(I,Z+1)n_\sd(I,Z+1)
-\sigma_{\sd\se}(I,Z)n_\sd(I,Z)],
\label{eq:evol_ndk}
\eeqn
respectively.
In equations \eqref{eq:evol_nik} and \eqref{eq:evol_nek}, $n_\sg^{(k)}$ and $n_\sg^{(k,l)}$
denote the number densities of neutrals ${\rm X}_\sg^{(k)}$ and ${\rm X}_\sg^{(k,l)}$.
We assume that neutral particles are much more abundant than charged particles and 
regard $n_\sg^{(k)}$ and $n_\sg^{(k,l)}$ as constant parameters.

In addition, we impose the charge neutrality condition
\beq
\sum_k n_\si^{(k)} - n_\se + \int dI \sum_Z Z n_\sd(I,Z) = 0.
\label{eq:neut}
\eeq
% It is easy to show using equations \eqref{eq:evol_nik}--\eqref{eq:evol_ndk}
% that the left-hand side of equation \eqref{eq:neut} is constant in time.

Equations \eqref{eq:evol_nik}--\eqref{eq:neut} form the closed set of basic equations 
for ionization-recombination reactions in gas-dust mixture.

%%%%%%%%%%%%%%%%%%%%%%%%%%%%%%%%%%%
\subsection{The equilibrium solution}
%%%%%%%%%%%%%%%%%%%%%%%%%%%%%%%%%%%
The equilibrium solution is obtained by imposing the conditions 
\beq
\dot{n}^{(k)}_\si = \dot{n}_\se = \dot{n}_\sd(I,Z) = 0
\label{eq:equil}
\eeq
for all $k$, $I$, and $Z$.

Usually, the equilibrium solutions are calculated 
with a reaction scheme involving many ion species and ion-neutral reactions
 (e.g., \citealt*{UN80,IN06}).
Because of the complexity of the ion-neutral reactions,
it is generally impossible to solve this problem analytically without any approximation or simplification.

This problem, however, becomes analytically tractable 
if we dot not require to distinguish ion species.
This is just achieved by introducing the total ion density
\beq
n_\si = \sum_k n_\si^{(k)}.
\label{eq:ni}
\eeq
Taking the sum of equations \eqref{eq:evol_nik}
over all $k$, we obtain the rate equation for $n_\si$, 
\beq
\dot{n}_\si = \zeta n_\sg 
- u_\si n_\si \dint dI \dsum_Z \sigma_{\sd\si}(I,Z)n_\sd(I,Z) - \beta n_\si n_\se,
\label{eq:evol_ni}
\eeq
where $n_\sg = \sum_k n_\sg^{(k)}$ is the total number density of the gas, and
\beq
u_\si = \frac{1}{n_\si}\sum_k n_\si^{(k)}u_\si^{(k)},
\label{eq:v}
\eeq
\beq
\beta = \frac{1}{n_\si}\sum_k n_\si^{(k)}\beta^{(k)},
\label{eq:alpha}
\eeq
\beq
\zeta = \frac{1}{n_\sg}\sum_k n_\sg^{(k)}\zeta^{(k)},
\label{eq:zeta}
\eeq
are  the average ion velocity, gas-phase recombination rate coefficient, and ionization rate, 
respectively.
Similarly, equations \eqref{eq:evol_nek}--\eqref{eq:neut} can be also written down in terms of 
$n_\si$, $u_\si$, $\beta$, and $\zeta$.
It is evident that equation \eqref{eq:evol_ni} is much simpler than the original 
equation \eqref{eq:evol_nik}.
This is mainly attributed to the cancellation of the last term in the original equation, 
i.e., the term describing the ion-neutral reactions.
Of course, this also means that we have lost the chance to know the composition of ions in detail.
This fact does not bother us since our primary interest is the charge state of dust aggregates,
not the composition of ions.

Now we try to solve the equations \eqref{eq:evol_nek}--\eqref{eq:neut} and \eqref{eq:evol_ni}
under the equilibrium condition \eqref{eq:equil} as analytically as possible.
First, equations \eqref{eq:evol_ni} and \eqref{eq:evol_nek}
are written as
\beq
\zeta n_\sg - u_\si\overline{\bracket{\sigma_{\sd\si}}}n_\sd n_\si - \beta n_\si n_\se = 0,
\label{eq:equil_xi}
\eeq
\beq
\zeta n_\sg - u_\se\overline{\bracket{\sigma_{\sd\se}}}n_\sd n_\se - \beta n_\si n_\se = 0,
\label{eq:equil_xe}
\eeq
respectively.
Here we have defined the averages of an arbitrary function $F=F(I,Z)$ over $Z$ and over $I$ as
\beq
\bracket{F}(I) \equiv \frac{1}{n_\sd(I)}\sum_Z F(I,Z)n_\sd(I,Z)
\label{eq:bracket_def}
\eeq
\beq
\overline{F}(Z) \equiv \frac{1}{n_\sd(Z)}\int F(I,Z)n_\sd(I,Z)dI
\label{eq:overline_def}
\eeq
with 
$n_\sd(I) \equiv \sum_Z n_\sd(I,Z)$ and $n_\sd(Z) \equiv \int n_\sd(I,Z)dI$, respectively.
We have also defined the total number density $n_\sd$ of dust aggregates by 
$n_\sd = \sum_Z \int n_\sd(I,Z) dI$.
Equations \eqref{eq:equil_xi} and \eqref{eq:equil_xe} can be easily solved 
in terms of $n_\si$ and $n_\se$ as
\beqn
&&n_\si = \frac{u_\se\overline{\bracket{\sigma_{\sd\se}}}n_\sd}{2\beta}
\left(\sqrt{1+\frac{4\beta\zeta n_\sg}{u_\si u_\se\overline{\bracket{\sigma_{\sd\si}}}\,\overline{\bracket{\sigma_{\sd\se}}}n_\sd^2}}-1\right), 
\label{eq:xi}
\\
&&n_\se = \frac{u_\si\overline{\bracket{\sigma_{\sd\si}}}n_\sd}{2\beta}
\left(\sqrt{1+\frac{4\beta\zeta n_\sg}{u_\si u_\se\overline{\bracket{\sigma_{\sd\si}}}\,\overline{\bracket{\sigma_{\sd\se}}}n_\sd^2}}-1\right).
\label{eq:xe}
\eeqn

Next, equation \eqref{eq:evol_ndk} is reduced to 
\beqn
&&u_\si n_\si\sigma_{\sd\si}(I,Z-1)n_\sd(I,Z-1)-u_\se n_\se\sigma_{\sd\se}(I,Z)n_\sd(I,Z)
\nonumber \\ 
&&=
u_\si n_\si\sigma_{\sd\si}(I,Z)n_\sd(I,Z) - u_\se n_\se\sigma_{\sd\se}(I,Z+1)n_\sd(I,Z+1).
\label{eq:equil_xd}
\eeqn
This equation means that the ``flux'' of the distribution $n_\sd(I,Z)$
from the state $Z-1$ to $Z$ must be 
balanced to that from $Z$ to $Z+1$.
If the flux did not vanish, there would exist a steady ``flow'' of the charge state distribution $n_\sd(I,Z)$
streaming from one direction to the other in $Z$-space.
However, since $n_\sd(I,Z)$ must be vanish at $Z\to\pm\infty$,
such a steady flow must not exist.
Therefore, both of the left and right-hand sides of equation~\eqref{eq:equil_xd}
must be zero. Hence we have
\beq
u_\si n_\si \sigma_{\sd\si}(I,Z)n_\sd(I,Z) = u_\se n_\se \sigma_{\sd\se}(I,Z+1)n_\sd(I,Z+1).
\label{eq:detailed_xd}
\eeq
This is just the condition of detailed balance between charge states $Z$ and $Z+1$.

For the sake of later convenience, we here rewrite the charge neutrality condition
\eqref{eq:neut} using the definitions \eqref{eq:ni}, \eqref{eq:bracket_def}, and
\eqref{eq:overline_def} as
\beq
n_\si - n_\se + \overline{\bracket{Z}}n_\sd = 0.
\label{eq:neut_x}
\eeq
As we will see later, this is the final equation that determines the equilibrium solution. 

The next step is to solve the detailed balance equation \eqref{eq:detailed_xd}.
To carry out the calculation, we need to specify the forms of the effective collision cross sections,
$\sigma_{\sd\si}$ and $\sigma_{\sd\se}$.
For simplicity, we model a fractal aggregate as a spherical, porous body
with radius $a$ and projected cross section $\sigma$.
Also, we neglect the electric polarization of aggregates \citep{DS87}
since a fractal aggregate is likely to have low dielectricity.
\citet{DS87} found that the strength of the electric polarization
relative to the charge-charge (Coulomb) interaction is determined by
a factor $(\epsilon-1)/(\epsilon +2)$, where $\epsilon$ is the dielectric constant
of the material.    
Given that a porous aggregate is well approximated by a dilute medium, 
the factor $(\epsilon-1)/(\epsilon +2)$ is inversely 
proportional to its mean internal density 
(Clausius-Mossotti relation; see, e.g., \citealt*{Jackson98}).
For example, the mean internal density of a fractal aggregate with $D\approx 2$ is
about inversely proportional to its radius $a$.
Therefore, we expect that the polarization effect of such a low-density aggregate is negligible.
Neglecting the polarization effect, 
the effective cross sections are simply written as \citep{Spitzer41}
\beqn
&&\sigma_{\sd\si}(I,Z) = \sigma s_\si(Z) \times\left\{
\begin{array}{ll}
\left(1-\dfrac{\lambda Z}{a}\right),  & Z < 0,  \\
\exp\left(-\dfrac{\lambda Z}{a}\right), \;\;  & Z \geq 0, 
\end{array}
\right.
\label{eq:sigma_di}
\\
&&\sigma_{\sd\se}(I,Z) =  \sigma s_\se(Z) \times\left\{
\begin{array}{ll}
 \exp\left(\dfrac{\lambda Z}{a}\right),\quad & Z < 0,  \\
 \left(1+\dfrac{\lambda Z}{a}\right),   & Z\geq 0, 
\end{array}
\right.
\label{eq:sigma_de}
\eeqn
where $s_{\si(\se)}(Z)$ is the probability that a colliding ion (electron) sticks to one of constituent monomers,
and $\lambda = e^2/\kB T$.
In this study, we assume that $s_\si(Z)$ and $s_\se(Z)$ are independent of the net charge $Z$
carried by an aggregate, i.e., $s_{\si(\se)}(Z) \equiv s_{\si(\se)}$.

As shown in Appendix, the solution $n_\sd(I,Z)$ to equation \eqref{eq:detailed_xd} 
with equations \eqref{eq:sigma_di} and \eqref{eq:sigma_de} 
is well approximated by a Gaussian distribution
\beq
n_\sd(I,Z) = \frac{n_\sd(I)}{\sqrt{2\pi\bracket{\Delta Z^2}_a}}
\exp\left[-\frac{(Z-\bracket{Z}_a)^2}{2\bracket{\Delta Z^2}_a}\right],
\label{eq:Gaussian}
\eeq
where
\beq
\bracket{Z}_a \equiv \frac{1}{n_\sd(I)}\sum_Z Zn_\sd(I,Z) 
\equiv -\frac{\Gamma a}{\lambda}
\label{eq:Zavr}
\eeq
and
\beq
\bracket{\Delta Z^2}_a \equiv \frac{1}{n_\sd(I)}\sum_Z(Z-\bracket{Z}_a)^2n_\sd(I,Z)
= \frac{1+\Gamma}{2+\Gamma} \frac{a}{\lambda}
\label{eq:Zvar}
\eeq
are the mean and dispersion of the charge distribution for fixed $a$, respectively.
This solution is valid when the radius $a$ is much larger than $\lambda$,
as is for aggregates much larger than constituent monomers (see Appendix).
The nondimensional parameter $\Gamma \equiv -\bracket{Z}\lambda/a = (-\bracket{Z}e^2/a)/\kB T$
measures the electrostatic attraction (repulsion) energy between
a charged aggregate and an incident ion (electron) relative to the thermal kinetic energy. 

Since the charge distribution is parametrized by $\Gamma$ only,
$n_\si$ and $n_\se$ can be written as a function of 
a single parameter $\Gamma$.
Using equations \eqref{eq:Gaussian}--\eqref{eq:Zvar},
$\bracket{\sigma_{\sd\si}}$ and $\bracket{\sigma_{\sd\se}}$ are
evaluated as 
\beq
\bracket{\sigma_{\sd\si}} = \sigma s_\si(1+\Gamma),
\label{eq:sigma_di_G}
\eeq
\beq
\bracket{\sigma_{\sd\se}} 
= \sigma s_\se \exp\left[-\Gamma +\frac{\lambda(1+\Gamma)}{2a(2+\Gamma)}\right] 
\approx \sigma s_\se \exp(-\Gamma),
\label{eq:sigma_de_G}
\eeq
respectively. 
Here we have used in equation \eqref{eq:sigma_de_G} that $a/\lambda \gg 1$.
Substituting equations \eqref{eq:sigma_di_G} and \eqref{eq:sigma_de_G}
into equations \eqref{eq:xi} and \eqref{eq:xe}, we have
\beq
n_\si = \frac{\zeta n_\sg}{s_\si u_\si \overline\sigma n_\sd}
\frac{\sqrt{1+2g(\Gamma)}-1}{(1+\Gamma)g(\Gamma)},
\label{eq:xi_G0}
\eeq
\beq
n_\se = \frac{\zeta n_\sg}{s_\se u_\se \overline\sigma n_\sd}
\frac{\sqrt{1+2g(\Gamma)}-1}{\exp(-\Gamma)g(\Gamma)},
\label{eq:xe_G0}
\eeq
where 
\beq
g(\Gamma) = 
\frac{2\beta\zeta n_\sg}{s_\si u_\si s_\se u_\se(\overline{\sigma}n_\sd)^2}\frac{\exp\Gamma}{1+\Gamma}.
\label{eq:g}
\eeq

Finally, the neutrality condition \eqref{eq:neut_x} with equations 
\eqref{eq:Zavr}, \eqref{eq:xi_G0}, and \eqref{eq:xe_G0} 
leads to the equation for $\Gamma$,
\beq
\frac{1}{1+\Gamma} - 
\left[\frac{s_\si u_\si}{s_\se u_\se}\exp\Gamma
 + \frac{1}{\Theta}\frac{\Gamma g(\Gamma)}{\sqrt{1+2g(\Gamma)}-1}\right] = 0,
\label{eq:neut_G0}
\eeq
where we have defined a nondimensional parameter
\beq
\Theta \equiv \frac{\zeta n_\sg \lambda}{s_\si u_\si \overline{\sigma}\,\overline{a} n_\sd^2}
= \frac{\zeta n_\sg e^2}{s_\si u_\si \overline{\sigma}\,\overline{a} n_\sd^2 \kB T}.
\label{eq:Theta}
\eeq
We have consequently arrived at the conclusion that
 all the conditions for ionization equilibrium, equations \eqref{eq:equil_xi}, \eqref{eq:equil_xe}, and \eqref{eq:detailed_xd}, are reduced to a single equation \eqref{eq:neut_G0} 
for a single parameter $\Gamma$.

To summarize the above analysis, 
we have considered the charge state of a dust-gas mixture in the presence of
ionization sources.
We have found that the self-consistent equilibrium solutions are
written as analytical functions of a single parameter $\Gamma$ (eqs. [\ref{eq:Gaussian}]--[\ref{eq:Zvar}],  [\ref{eq:xi_G0}], and [\ref{eq:xe_G0}]), and have obtained the equation for this master parameter 
(eq. [\ref{eq:neut_G0}]).
This equation can be easily solved numerically, and thus gives semianalytical solutions to
 $n_\si$, $n_\se$, and $n_\sd(I,Z)$.

The resultant equations can be further simplified
when the gas-phase recombination rate coefficient $\beta$ is so small that 
the factor $g$ defined in equation \eqref{eq:g} is much less than unity.
In this limit, equations \eqref{eq:xi_G0}, \eqref{eq:xe_G0}, and \eqref{eq:neut_G0}
are simply rewritten as
\beq
n_\si = \frac{\zeta n_\sg}{s_\si u_\si \overline\sigma n_\sd}\frac{1}{1+\Gamma},
\label{eq:xi_G}
\eeq
\beq
n_\se = \frac{\zeta n_\sg}{s_\se u_\se \overline\sigma n_\sd}\exp\Gamma,
\label{eq:xe_G}
\eeq
and
\beq
\frac{1}{1+\Gamma} -\left[ \frac{s_\si u_\si}{s_\se u_\se}\exp\Gamma + \frac{\Gamma}{\Theta}\right] = 0,
\label{eq:neut_G}
\eeq
respectively. 
As seen in \S3.1.3, this approximation is valid in typical protoplanetary disks 
unless the dust is significantly depleted (e.g., by vertical sedimentation). 

%%%%%%%%%%%%%%%%%%%%%%%%%%%
\subsection{Limiting cases}
%%%%%%%%%%%%%%%%%%%%%%%%%%%
Equation \eqref{eq:neut_G0}, or \eqref{eq:neut_G}, provides clear insight into
the charge state of a gas-dust mixture.
The first and second terms in the bracket in this equation
originate from $n_\se$ and $-\bracket{Z}n_\sd$ in the charge neutrality
condition \eqref{eq:neut}, respectively.
This means that the parameter $\Theta$ determines which of free electrons and dust aggregates 
are the dominant carriers of negative charge.
In the following, we categorize the charge state from limiting cases of equation \eqref{eq:neut_G}.

%%%%%%%%%%%%%%%%%%%%%%%%%%%%%%%%%%%%%%%%%%%%%%%%%%%%%%%%%%%%%
\subsubsection{$\Theta \to \infty$: ion-electron plasma limit}
%%%%%%%%%%%%%%%%%%%%%%%%%%%%%%%%%%%%%%%%%%%%%%%%%%%%%%%%%%%%%
In the limit $\Theta \to \infty$, the contribution from charged dust becomes negligibly small.
Hence, equation \eqref{eq:neut_G} is well approximated by 
\beq
\frac{1}{1+\Gamma} \approx \frac{s_\si u_\si}{s_\se u_\se}\exp\Gamma.
\label{eq:ion-elec}
\eeq
This is a well-known charge-equilibrium condition for
a dust particle immersed in an ordinary ion-electron plasma \citep{Spitzer41,SM02}.
We denote the solution $\Gamma$ to equation \eqref{eq:ion-elec} by $\Gamma_{\rm max}$, 
since this is the maximum value of $\Gamma$ obtained from equation \eqref{eq:neut_G}.
Typically, $\Gamma_{\rm max}$ takes a value of order unity.

%%%%%%%%%%%%%%%%%%%%%%%%%%%%%%%%%%%%%%%%%%%%%%%%%%%%%
\subsubsection{$\Theta \to 0$: ion-dust plasma limit}
%%%%%%%%%%%%%%%%%%%%%%%%%%%%%%%%%%%%%%%%%%%%%%%%%%%%%
In the limit $\Theta \to 0$, on the other hand, the contribution from free electrons 
becomes negligible.
This means that the dominant carriers of negative charges are dust particles, not free electrons.
We shall refer to this limit as the {\it ion-dust plasma} limit.
In this limit, equation \eqref{eq:neut_G} can be approximated by
\beq
\frac{1}{1+\Gamma} \approx \frac{\Gamma}{\Theta}.
\eeq
The solution is easily obtained as
\beq
\Gamma \approx \frac{\sqrt{1+4\Theta^2} -1}{2}
\approx \Theta.
\label{eq:ion-dust}
\eeq
Since $\Gamma$ is now negligibly small, the effective cross sections $\bracket{\sigma_{\sd\si(\se)}}$
 are approximately equal to $\sigma s_{\si(\se)}$. 
Equations \eqref{eq:xi_G} and \eqref{eq:xe_G}
lead to the ratio of $n_\se$ to $n_\si$,
\beq
\frac{n_\se}{n_\si} \approx \frac{s_\si u_\si}{s_\se u_\se} 
\approx \frac{s_\si}{s_\se}\sqrt{\frac{m_\se}{m_\si}}.
\label{eq:xe/xi}
\eeq
where we have used that $u_{\rm i(e)} = \sqrt{8\kB T/\pi m_{\rm i(e)}}$.
We note that this value of $n_\se/n_\si$ is larger than that of \citet{Umebayashi83} by a factor of 
$1/\sqrt{s_\se}$.
This is because we have assumed the value of $s_\se$ as independent of $Z$ while \citet{Umebayashi83}
 considered $s_\se(Z>0)=1$.
As far as the author knows, there is no experimental data that validates either of the assumptions.
However, this difference is practically unimportant unless the ratio
$\sqrt{s_\se(Z<0)/s_\se(Z\geq0)}$ is much less than unity.

The value of  $\Theta$ at the transition from one plasma regime to the other
can be roughly estimated by equating the asymptotic solutions for both limits,
 i.e., $\Theta \approx \Gamma_{\rm max}$.
We will use this estimation in the next section.

%%%%%%%%%%%%%%%%%%%%%%
\section{Application: electric barrier against dust growth}
%%%%%%%%%%%%%%%%%%%%%%
Electrostatic interaction between charged aggregates
affect their collisional cross section.
Let us consider two dust aggregates with mass $m_j$, radius $a_j$ and charge $Z_je$,
where $j(=1,2)$ labels the aggregates.
The kinetic energy for relative motion of two aggregates 1 and 2 is
written as 
\beq
E_{\rm kin} = \frac{1}{2} \tilde{m} (\Delta u)^2, %= \frac{1}{4} m  (\Delta v)^2
\label{eq:Ekin0}
\eeq 
where $\tilde{m}=m_1m_2/(m_1+m_2)$ and $\Delta u$ are the reduced mass and the relative speed
for the aggregate pair.
The electrostatic energy between the aggregates just before contact is
\beq
E_{\rm el} = \frac{Z_1Z_2e^2}{a_1+a_2}. %= \frac{Z_1Z_2e^2}{2a} 
\label{eq:Eel0}
\eeq
Neglecting the polarization effect as done in \S2,
the effective collision cross section $\sigma_{\rm dd}$ for the aggregates 
is expressed as \citep{Spitzer41}
\beq
\sigma_{\rm dd} = \left\{
\begin{array}{ll}
\pi(a_1+a_2)^2 \left(1-\dfrac{E_{\rm el}}{E_{\rm kin}}\right), & E_{\rm kin} > E_{\rm el},  \\
0, & E_{\rm kin} \leq E_{\rm el} , 
\end{array}
\right.
\eeq
Therefore, the condition for the aggregates to collide with each other is
\beq
E_{\rm kin} > E_{\rm el}.
\label{eq:growthcond}
\eeq
In this section, we examine whether this condition is satisfied in an early stage of 
dust evolution in a protoplanetary disk.

%%%%%%%%%%%%%%%%%%%%%%%
\subsection{Model setup}
%%%%%%%%%%%%%%%%%%%%%%%
%%%%%%%%%%%%%%%%%%%%%%%%%%%
\subsubsection{Protoplanetary disk model}
%%%%%%%%%%%%%%%%%%%%%%%%%%%
We assume that the gas surface density $\Sigma_\sg$ and the temperature $T$ of the disk 
obey power laws
\beq
\Sigma_\sg(r) = 1.7\times 10^3 f_\Sigma \pfrac{r}{1\,\AU}^{-3/2} {\rm g/cm^2},
\label{eq:Sigma} 
\eeq
and
\beq
T(r) = 280\pfrac{r}{1 \,\AU}^{-1/2} {\rm K},
\label{eq:T}
\eeq
where $r$ is the distance from the central star and $f_\Sigma$ is a nondimensional scaling parameter.
The model with $f_\Sigma=1$ is known as the minimum-mass solar nebula (MMSN) model \citep{Hayashi81}.
We adopt $f_\Sigma = 1$ unless otherwise noted.
The temperature profile \eqref{eq:T} is valid only for optically thin regions.
We nevertheless employ this profile throughout the disk since
the main result is insensitive to the detail of the temperature profile
(see eq.~[\ref{eq:Eratio}] below).

The hydrostatic equilibrium in the vertical direction gives 
the gas density distribution
\beq
\rho_\sg(r,z) = \frac{\Sigma_\sg}{\sqrt{2\pi}H(r)}\exp\left[-\frac{z^2}{2H(r)^2} \right],
\eeq
where $z$ is the height from the disk midplane and 
$H(r)=c_s(r)/\Omega_{\rm K}(r)$ is the gas scale height.
The isothermal sound velocity $c_s$ and the Kepler rotational frequency $\Omega_{\rm K}(r)$
are given by $c_s(r) = \sqrt{\kB T(r)/\mu m_{\rm H}}$ and $\Omega_{\rm K}(r) = \sqrt{GM_*/r^3}$,
where $\mu$ is the mean molecular weight, $m_{\rm H}$ is the hydrogen mass, 
$G$ is the gravitational constant, and $M_*$ is the mass of the central star.
We adopt $\mu=2.34$ and $M_*=1M_\sun$ in the following calculation. 
The total number density $n_{\sg}$ of gas particles is given by 
$n_{\sg} = \rho_{\sg}/\mu m_{\rm H}$.

We assume that  dust material is well mixed in the disk and that 
the dust density $\rho_\sd$ is simply related to the gas density $\rho_\sg$ by
\beq
\rho_\sd(r,z) = f_{\sd\sg}\rho_\sg(r,z),
\eeq
where $f_{\sd\sg}$ is the dust-to-gas ratio in the disk.
The solar system abundance of condensates including water ice estimated by \citet{Pollack+94}
leads to the dust-to-gas ratio $f_{\sd\sg} = 0.014$ as well as
the monomer bulk density $\rho_0 = 1.4{\rm g/cm^3}$ \citep{THI05}.
We do not consider the sublimation of water ice in inner disk regions for simplicity.

%%%%%%%%%%%%%%%%%%%%%%%%%%%%%%
\subsubsection{Ionization rate}
%%%%%%%%%%%%%%%%%%%%%%%%%%%%%%
In this study, we consider Galactic cosmic rays \citep{UN81}, stellar X-rays \citep{IG99},
and radionuclides \citep{UN09} as the ionizing sources.
Thus, we decompose the ionization rate $\zeta$ as
\beq
\zeta \approx \zeta_{\rm CR} + \zeta_{\rm XR} + \zeta_{\rm RA}
\eeq
where $\zeta_{\rm CR}$, $\zeta_{\rm XR}$, and $\zeta_{\rm RA}$ denote the rate of ionization by 
cosmic rays, X-rays, and radionuclides, respectively.
We do not consider thermal ionization. This is negligible for $T\ll 10^3{\rm K}$, or for $r \gg 0.1\AU$ \citep{Umebayashi83}.
Charged particles are created primarily by ionization of ${\rm H_2}$ and ${\rm He}$.
The ionization rate for ${\rm He}$ is related to that for ${\rm H_2}$ 
by $\zeta^{\rm (He)} = 0.84\zeta^{\rm (H_2)}$ \citep{UN90,UN09}, 
so it is sufficient to know $\zeta^{\rm (H_2)}$ only.
The total ionization rate $\zeta$ is given by 
$\zeta = \zeta^{\rm (H_2)}x_{\rm H_2} + \zeta^{\rm (He)}x_{\rm He}$,
where $x_{\rm H_2} = n_{\rm H_2}/n_\sg$ and $x_{\rm He} = n_{\rm He}/n_\sg$
are the fractional abundances of ${\rm H_2}$ and ${\rm He}$. 
We calculate $x_{\rm H_2}$ and $x_{\rm He}$ from the solar system abundance by \citet{AG89}. 

The cosmic-ray ionization rate $\zeta^{(\rm H_2)}_{\rm CR}$ for ${\rm H_2}$ is 
 given by a fitting formula \citep{UN09}
\beqn
\zeta^{(\rm H_2)}_{\rm CR}(r,z) &\approx&  \frac{\zeta^{\rm (H_2)}_{\rm CR,0}}{2}
\left\{ \exp\left(-\dfrac{\chi^+_\sg(r,z)}{\chi_{\rm CR}}\right)
\left[1+\pfrac{\chi^+_\sg(r,z)}{\chi_{\rm CR}}^{3/4}\right]^{-4/3} \right. \nonumber \\
&&+ \left. \exp\left(-\dfrac{\chi^-_\sg(r,z)}{\chi_{\rm CR}}\right)
\left[1+\pfrac{\chi^-_\sg(r,z)}{\chi_{\rm CR}}^{3/4}\right]^{-4/3} \right\},
\label{eq:zetaCR}
\eeqn
where $\zeta^{\rm(H_2)}_{\rm CR,0} \approx 1.0\times 10^{-17}{\rm/s}$ 
is the cosmic-ray ionization rate for ${\rm H_2}$ in the interstellar space, 
$\chi_{\rm CR} \approx 96\,{\rm g/cm^2}$ is the attenuation length of the ionization rate,
and $\chi^+_\sg(r,z) = \int_z^\infty\rho_\sg(r,z')dz'$ and $\chi^-_\sg(r,z) = \Sigma_\sg(r) -\chi^+_\sg(r,z)$ are the vertical gas column densities measured from the upper and lower infinities,
respectively.
For the radionuclide ionization rate, we assume $\zeta^{\rm (H_2)}_{\rm RA} \approx 7 \times 10^{-19}{\rm /s}$, which corresponds to the ionization rate by
${\rm ^{26}Al}$ with an abundance ratio  ${\rm ^{26}Al/^{27}Al} = 5 \times 10^{-5}$ \citep{UN09}.
% We do not consider stellar X-rays \citep{GNI97,IG99} as additional 
% ionizing sources, because the X-ray ionization is important only at small vertical depths ($\chi^+_\sg,\chi^-_\sg \la 10{\rm g/cm^2}$) and thus at high altitudes ($z \ga 2H$ at $r=1{\rm AU}$).

The stellar X-ray ionization rate has been calculated by \citet{IG99} 
using the Monte Carlo radiative transfer code including Compton scattering. A useful fitting formula is given by \citet{TS08},
\beqn
\zeta_{\rm XR}(r,z) &\approx& \zeta_{\rm XR,0}
\pfrac{r}{1\AU}^{-2}\pfrac{L_{\rm XR}}{2\times 10^{30}{\rm erg/s}} \nonumber \\
&&\times
\left\{ \exp\left(-\dfrac{\chi^+_\sg(r,z)}{\chi_{\rm XR}}\right)
%\left[1+\pfrac{\chi^+_\sg(r,z)}{\chi_{\rm XR}}^{3/4}\right]^{-4/3} 
%\right. \nonumber \\
+ \exp\left(-\dfrac{\chi^-_\sg(r,z)}{\chi_{\rm XR}}\right)
%\left[1+\pfrac{\chi^-_\sg(r,z)}{\chi_{\rm XR}}^{3/4}\right]^{-4/3} 
\right\}, 
\label{eq:zetaXR}
\eeqn
where $L_{\rm XR}$ is the X-ray luminosity, and 
$\zeta_{\rm XR,0} = 2.6\times 10^{-15}{\rm /s}$ and $\chi_{\rm XR} = 8.0 {\rm g/cm^2}$ 
are the fitting parameters.
%Comparing this formula with figure 5 of \citet{IG99},
%we find that the parameter set of $\zeta_{\rm XR,0}\approx6.9 \times 10^{-16}{\rm /s}$ 
%and $\chi_{\rm XR} \approx 8.5{\rm g/cm^2}$ 
This fitting formula approximately reproduces the $\kB T_{\rm XR} = 5{\rm keV}$ result of \citet{IG99}
within the column density range $\chi_\sg \ga 1{\rm g/cm^2}$,
where scattered hard ($\ga 5{\rm keV}$) X-rays are responsible for the ionization.
We use equation \eqref{eq:zetaXR} in the following calculation 
since the typical value of $\chi_\sg$ is within the above range.
% A similar fitting formula was obtained by \citet{TS08},
% but our formula agrees with the result of \citet{IG99} better
% for $\chi_\sg \approx 1{\rm-}10\,{\rm g/cm^2}$, 
% for which the X-ray ionization dominates over the other ionization processes.
We take $L_{\rm XR} = 2\times 10^{30}{\rm erg/s}$ in accordance with the median characteristic
X-ray luminosity observed by {\it Chandra} for young solar-mass stars in the Orion Nebula Cluster \citep{Wolk+05}.
Although the characteristic X-ray temperature $\kB T_{\rm XR} \approx 2.4{\rm keV}$ observed by \citet{Wolk+05} is lower than 
the assumed value of $\kB T_{\rm XR} \approx 5 {\rm keV}$, 
the choice of the temperature does not significantly affect the resulting ionization rate \citep{IG99}. 
We do not consider temporary increase in the X-ray luminosity 
due to stellar flaring \citep{Wolk+05}, since our analytical method assumes stationary ionization processes. 
As pointed out by \citet{IN06c}, the flaring could quantitatively change the ionization state of the disk.
We will examine the effect of the time-dependent flaring on dust growth in the future work.
  
\begin{figure*}
\plottwo{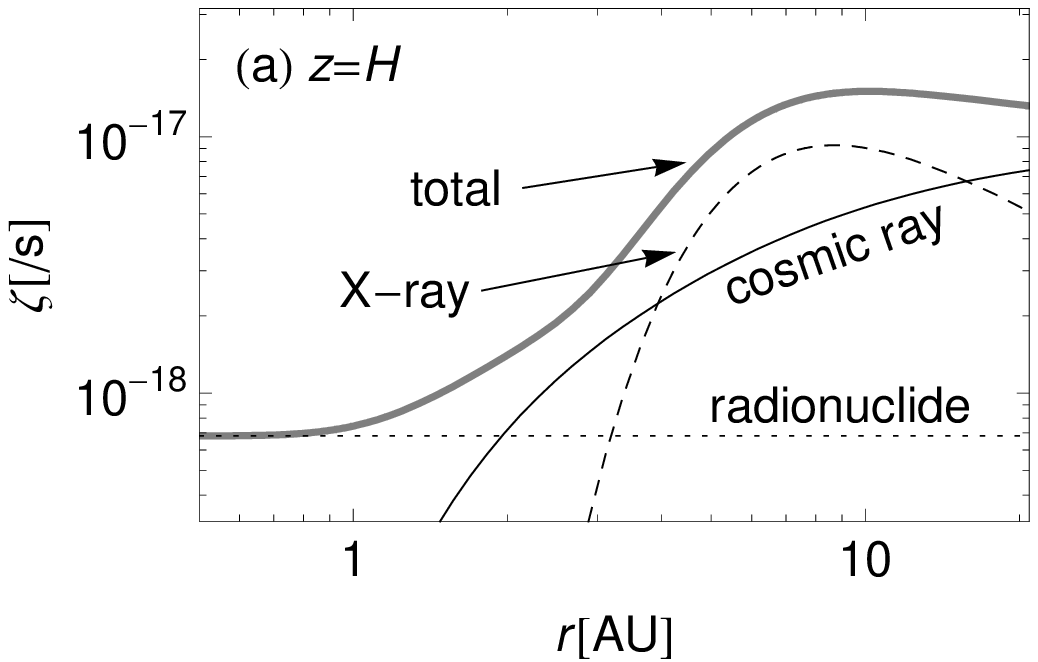}{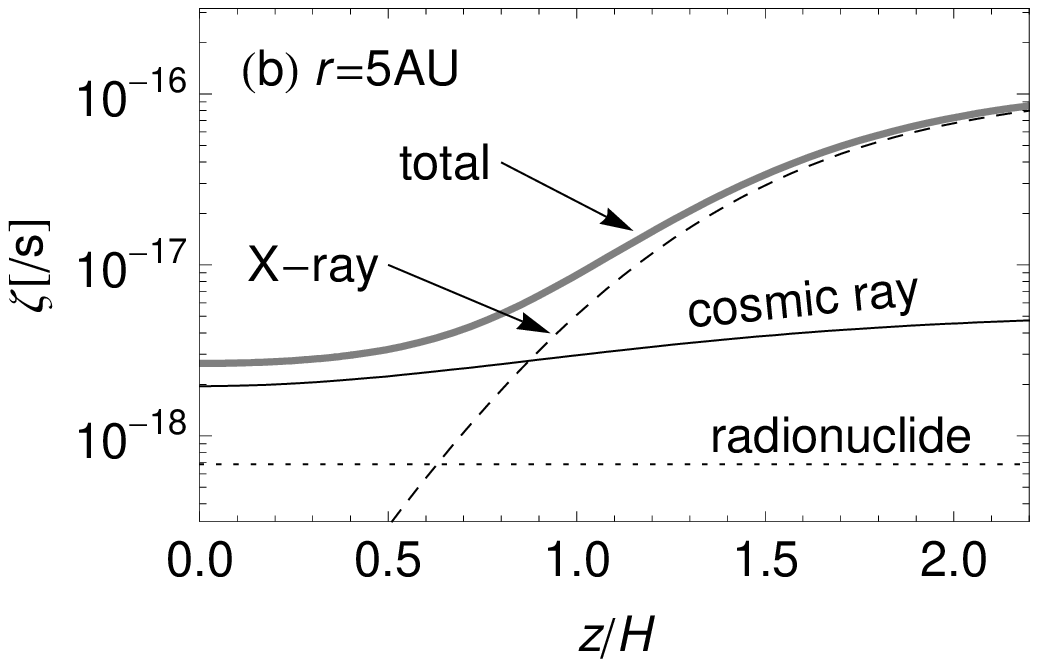}
\caption{Total ionization rate $\zeta$ ({\it thick gray curves})
at different disk radii ({\it a}) and and altitudes ({\it b}).
Here $H$ denotes the scale height of the disk. The solid, dashed, and dotted black curves represent the contribution from Galactic cosmic rays, stellar X-rays, and radionuclides, respectively.}
\label{fig:zeta}
\end{figure*}

Figure \ref{fig:zeta} shows the total ionization rate $\zeta$ as well as 
its three components ($\zeta_{\rm CR}$, $\zeta_{\rm XR}$, and $\zeta_{\rm RA}$) 
as a function of $r$ and $z$. 
X-ray ionization is dominant at outer radii $(r\ga 4{\rm AU})$, while
radionuclide ionization dominates at inner radii $(r\la 1{\rm AU})$.
Cosmic-ray ionization is important in outer $(r\ga 2{\rm AU})$ and low-altitude $(z \la H$) regions.

%%%%%%%%%%%%%%
%%%%%%%%%%%

%%%%%%%%%%%%%%%%%%%%%%%%%%%%%%
\subsubsection{Dust growth model}
%%%%%%%%%%%%%%%%%%%%%%%%%%%%%%
Based on the results of recent laboratory and computer simulations mentioned in \S1,
we model the dust growth in protoplanetary disks as follows.
We start with monodisperse, non-aggregated dust grains (monomers) with radius $a_0$.
The mass $m_0$ and number density $n_{\sd0}$ of monomers are then written as
$m_0=(4\pi/3)\rho_0a_0^3$ and $n_{\rm d0} = \rho_\sd/m_0$.
The dust is assumed to grow into an ensemble of quasi-monodisperse, fractal aggregates
with typical monomer number $N$ and fractal dimension $D \sim 2$.
Under this assumption, we may regard $N$ as the label of dust evolutionary stages.
Theoretically, this type of growth is the best modeled by the so-called {\it ballistic cluster-cluster
aggregation} (BCCA; e.g., \citealt*{Meakin91}).
For this reason, we shall refer to the quasi-monodisperse fractal growth 
as the ``BCCA growth.''\footnote{Exactly speaking, the BCCA (i.e., collision between {\it identical} aggregates; see \citealt*{Meakin91})
can only occur when the relative velocity of aggregates is induced by Brownian motion.
When the relative velocity is induced by {\it differential} sedimentation,
 any {\it identical} aggregates cannot collide with each other, 
and therefore a BCCA cluster in its original sense cannot be created.
On the other hand, a laboratory experiment by \citet{Blum+98} shows that the outcome of sedimentation-driven coagulation is an ensemble of quasi-monodisperse, fractal aggregates with 
$D \approx 1.7$,
as is for Brownian-motion-driven growth (e.g., \citealt*{Kempf+99}).
This may be explained by the fact that the dominant growth mode in differential sedimentation is
the collision between {\it similar} aggregates \citep{THI05}.
For this reason, this study treats the growth by differential sedimentation as the ``BCCA growth.''
}
Note that the number density $n_\sd$ of aggregates is inversely proportional to $N$, 
since $n_{\rm d0} = Nn_\sd$ is conserved.
The fractal growth continues until colliding aggregates
become energetic enough to compress each other.
According to the microscopic model of \citet{DT97},
the critical kinetic energy for the onset of collisional compression 
is given by $E_{\rm kin} \sim E_{\rm roll}$, where
\beqn
E_{\rm roll} &=& 6\pi^2 \gamma \frac{a_0}{2}\xi_{\rm crit} \nonumber \\
&\approx& 5.9 \times 10^{-10} \pfrac{\gamma}{100{\rm erg/cm^2}}
\pfrac{\xi_{\rm crit}}{2{\rm\mathring{A}}}\pfrac{a_0}{0.1\micron} {\rm erg}\qquad
\label{eq:Eroll}
\eeqn
is the energy needed to roll a monomer on another monomer in contact by $90^\circ$,
$\gamma$ is the surface adhesion energy for the two monomers,
and $\xi_{\rm crit}$ is the critical tangential displacement for starting the rolling.
For icy monomers, $\gamma$ is estimated as $\gamma \approx 100{\rm g/cm^2}$ \citep{I92} 
but a realistic value of $\xi_{\rm crit}$ is unknown.
For a conservative estimation, we assume the minimum displacement 
$\xi_{\rm crit} = 2{\rm\mathring{A}}$ anticipated by the theory \citep{DT97}, which makes our aggregates the most easily compressed.
The assumed value of $E_{\rm roll}$ is not much different from
 the experimental value for rocky (${\rm SiO_2}$) monomers,
$E_{\rm roll} \approx 1.3\times 10^{-9}(a_0/0.1\micron){\rm erg}$ 
 \citep{Heim+99}, so the duration of the fractal growth stage
is insensitive to our choice of dust material.
We restrict our calculation to an early growth stage
where the relative kinetic energy $E_{\rm kin}$ does not exceed 
the critical rolling energy $E_{\rm roll}$.

The radius $a$ of a fractal aggregate is approximately given by
\beq
a \approx a_0 N^{1/D}.
\eeq
A classical, compact aggregate has $D=3$, while a BCCA cluster has $D \approx 1.9$ \citep{Meakin91}.
We adopt $D=2$ in the following calculation.
The projected cross section $\sigma$ of an aggregate is simply set to 
$\sigma \approx \pi a^2 \approx \sigma_0 N^{2/D} \approx \sigma_0N $,
where $\sigma_0 = \pi a_0^2$ is the geometrical cross section of a monomer.
This assumption is consistent with the fact that $\sigma \propto N$ for $D\la 2$ \citep{MD88,MDM89,Minato+06}.
Note that the quantity $\sigma n_\sd$ is independent of $N$, i.e., conserved for the BCCA growth.

Assuming a quasi-monodisperse size distribution, the kinetic energy \eqref{eq:Ekin0} is written as
\beq
E_{\rm kin} \approx \frac{1}{4}m(\Delta u)^2,
\label{eq:Ekin}
\eeq
where we have used $\tilde{m} \approx m/2$.
In a protoplanetary disk, relative motion of aggregates is induced by Brownian motion,
sedimentation toward the midplane of the disk, and turbulence.
We therefore write the relative velocity $\Delta u$ as
\beq
\Delta u \approx \sqrt{(\Delta u_{\rm Brown})^2 + (\Delta u_{\rm sed})^2 +(\Delta u_{\rm turb})^2},
\eeq
where $\Delta u_{\rm Brown}$, $\Delta u_{\rm sed}$, and $\Delta u_{\rm turb}$ are the 
relative speed induced by the Brownian motion, differential sedimentation, and turbulence, 
respectively.

The mean relative speed of the Brownian motion is given by
\beq
\Delta u_{\rm Brown} = \sqrt{\frac{8\kB T}{\pi\tilde{m}}} \approx \sqrt{\frac{16\kB T}{\pi m}}.
\eeq
In fact, the relative speed of the Brownian motion fluctuates according to the Maxwell distribution,
and aggregates have a chance to get a relative kinetic energy $E$ much larger than the thermal energy 
$\sim \kB T$ with a probability $\propto E^{1/2} \exp(-E/\kB T)$. 
However, as we see later, the effect of the thermal velocity fluctuation is insignificant,
since the electrostatic energy can go up to $10^5\kB T$. 

The relative speed induced by the differential sedimentation is
\beq
\Delta u_{\rm sed} = \Omega_{\rm K}^2z\Delta t_{\rm stop},
\label{eq:vsed}
\eeq
where $t_{\rm stop}$ is the stopping time of an aggregate.
For aggregates smaller than the mean free path $\ell_\sg$ of gas particles, 
$t_{\rm stop}$ is given by Epstein's law 
\beq
t_{\rm stop} = \frac{3}{4\rho_\sg u_\sg}\frac{m}{\sigma},
\label{eq:tstop}
\eeq
where $u_\sg = \sqrt{8\kB T/\pi\mu m_{\rm H}}$ is the mean thermal speed of gas particles.
The mean free path in our disk model is calculated to be $\ell_\sg \sim 1 (r/1\,\AU)^{11/4}\,{\rm cm}$,
which is much larger than a typical size of aggregate which we are interested in.
For this reason, we always use Epstein's law \eqref{eq:tstop} in the following calculation. 
Also, we replace $\Delta t_{\rm stop}$ with its maximum value $t_{\rm stop}$,
which leads to the most conservative evaluation of dust charging effect.

The relative velocity induced by turbulence is given by 
\beq
\Delta u_{\rm turb} \approx \frac{u_{\rm small}}{t_{\rm small}}\Delta t_{\rm stop},
\label{eq:vturb}
\eeq
where $u_{\rm small}$ and $t_{\rm small}$ are the characteristic velocity
and turnover time of the smallest turbulent eddies, respectively
\citep{Weidenschilling84,OC07}.
This expression is valid for aggregates with stopping times $t_{\rm stop}$ much smaller
than $t_{\rm small}$.
The velocity and turnover time of the largest eddies, $u_{\rm large}$ and $t_{\rm large}$,
are related to $u_{\rm small}$ and $t_{\rm small}$ by 
$u_{\rm large} = {\rm Re}^{1/4}u_{\rm small}$ and $t_{\rm large} \approx {\rm Re}^{1/2}t_{\rm small}$,
where ${\rm Re = \nu_{\rm turb}/\nu_{\rm mol}}$ is the Reynolds number.
The molecular viscosity $\nu_{\rm mol}$ is written as $\nu_{\rm mol} = 0.5u_\sg/n_\sg\sigma_{\rm mol}$,
where $\sigma_{\rm mol}=2\times10^{-15}{\rm cm^2}$ is the molecular collision cross section
\citep{CC70}.
We express the turbulence viscosity $\nu_{\rm turb}=u_{\rm large}^2t_{\rm large}$ as
$\nu_{\rm turb} = \alpha_{\rm turb}c_s^2\Omega_{\rm K}^{-1}$, 
where $\alpha_{\rm turb}$ is the $\alpha$-parameter.
The turnover time of the largest eddies is taken to be $t_{\rm large} \approx \Omega_{\rm K}^{-1}$,
and thus their velocity is given by $u_{\rm large} \approx \sqrt{\alpha_{\rm turb}}c_s$.
We consider $\alpha_{\rm turb} = 0,\, 10^{-4},10^{-3},$ and $10^{-2}$ in this study.
Again, we replace $\Delta t_{\rm stop}$ in equation \eqref{eq:vturb} with $t_{\rm stop}$.

If both of colliding aggregates have the mean charge $\bracket{Z}$, 
the electrostatic energy \eqref{eq:Eel0} is written as
\beq
E_{\rm el} \approx \frac{\bracket{Z}^2e^2}{2a} = \frac{\Gamma^2 a}{2\lambda^2}.
\label{eq:Eel}
\eeq
where $\Gamma$ is the master parameter defined in equation \eqref{eq:Zavr}.
Equation \eqref{eq:Eel} overestimates a true repulsion energy 
when one (or both) of the aggregates has a positive charge $Z>0$, or a negative charge $-Z>0$ smaller than the average value $-\bracket{Z}$.
To account for the dispersion of the charge distribution,
we introduce the ``three-sigma'' electrostatic energy
\beq
E_{\rm el,3\sigma} \equiv \frac{\bracket{Z}(\bracket{Z}+3\bracket{\Delta Z^2}^{1/2})e^2}{2a},
\label{eq:Eel_3sigma}
\eeq
where the dispersion $\bracket{\Delta Z^2}^{1/2}$ is calculated from equation \eqref{eq:Zvar}.
$E_{\rm el,3\sigma}$  represents the electrostatic energy 
between two aggregates with charges $\bracket{Z}$ and 
$\bracket{Z}+3\bracket{\Delta Z^2}^{1/2}$. 
The probability that a collision involves the electrostatic energy larger than $E_{\rm el,3\sigma}$
is as small as ``three sigma'' ($\sim 10^{-3}$).
Note that $E_{\rm el,3\sigma}$ is always smaller than $E_{\rm el}$ 
since $\bracket{Z}$ is always negative.
Also, $E_{\rm el,3\sigma}$ becomes negative if $-\bracket{Z}$
 is smaller than $3\bracket{\Delta Z^2}^{1/2}$.
This actually happens when the aggregate size is sufficiently small (see fig.~\ref{fig:Zx}{\it a} below).

We compute the charge state at each evolutionary stage from equation \eqref{eq:neut_G}
with $\overline{a}\approx a(N)$, $\overline{\sigma} \approx \sigma(N)$.
The sticking coefficients are estimated by phonon theory to be $s_\si \approx 1$ and $s_\se=0.1 \,\dots\, 1$ \citep{Umebayashi83}.
We adopt $s_\si = 1$ and $s_\se = 0.3$ in this study.
The results obtained in this section are insensitive to the choice of $s_\se$ 
as long as $0.1\la s_\se \la 1$.
We do not use the original equation \eqref{eq:neut_G0} because the gas-phase ionization 
is negligible in the present case.
The gas-phase recombination rate is typically $\beta \sim 10^{-12\,\dots\,-7}\,{\rm cm^3/s}$.  
Using this value, we can estimate $g(\Gamma)$ as
\beqn
g(\Gamma)&\sim& \frac{\beta\zeta n_\sg}{u_\si s_\se u_\se(\sigma n_\sd)^2} \nonumber \\
 &\sim&  10^{-11\,\dots\,-6} \pfrac{f_{\sd\sg}}{10^{-2}}^{-2} \pfrac{r}{5\AU}^{3}\pfrac{T}{130{\rm K}}^{-1/2}
\pfrac{\zeta}{10^{-17}{\rm /s}}, \nonumber \\
\eeqn
independently of $N$. Therefore, equation \eqref{eq:neut_G} is valid unless
dust is depleted and $f_{\sd\sg}$ decreases by many orders of magnitude.

To confirm that our semianalytical calculation does work well,
we have also performed fully numerical calculations including multi-component ions.
In the numerical calculations, a simple reaction model by \citet{UN80} is adopted.
This reaction model involves five light ions (${\rm H^+, H_2^+, H_3^+, He^+, C^+}$),
heavy molecular ions (${\rm m^+}$), metal ions (${\rm M^+}$), free electrons, and charged dust aggregates.
Heavy molecular ions and metal ions are represented by ${\rm HCO^+}$ and ${\rm Mg^+}$, 
respectively.
We adopt the same values of the rate coefficients $\beta^{(k)}$, $\beta^{'(k,l)}$ 
and the neutral gas abundances $n_\sg^{(k,l)}/n_\sg$ as those used by \citet{Sano+00}.
As seen below, the dominant ions are metal ions, 
which is essentially due to the fast charge transfer from heavy molecules to
metal atoms. 
With this fact, we set the average ion mass $m_\si$ to be the mass of ${\rm Mg^+}$
($m_\si = 24m_{\rm H}$) in semianalytical calculations. 
The numerical solutions are obtained from equations \eqref{eq:evol_nik}--\eqref{eq:equil}.
We remark that the full numerical calculation is far more time-consuming than
the semianalytical one.

%%%%%%%%%%%%%%%%%%%%
\subsection{Results}
%%%%%%%%%%%%%%%%%%%%
%%%%%%%%%%%%%%%%%%%%%%%%%%%%%%%%%%%%%%%%%%%%%%%%%%%%%%%%%%%%%%%%%%%%%%%%%%%%%
\subsubsection{The fiducial case: $\alpha_{\rm turb} = 0,\,a_0 = 0.1\micron$}
%%%%%%%%%%%%%%%%%%%%%%%%%%%%%%%%%%%%%%%%%%%%%%%%%%%%%%%%%%%%%%%%%%%%%%%%%%%%%
%%%%%%%%%%%%%%%
\begin{figure*}
\plottwo{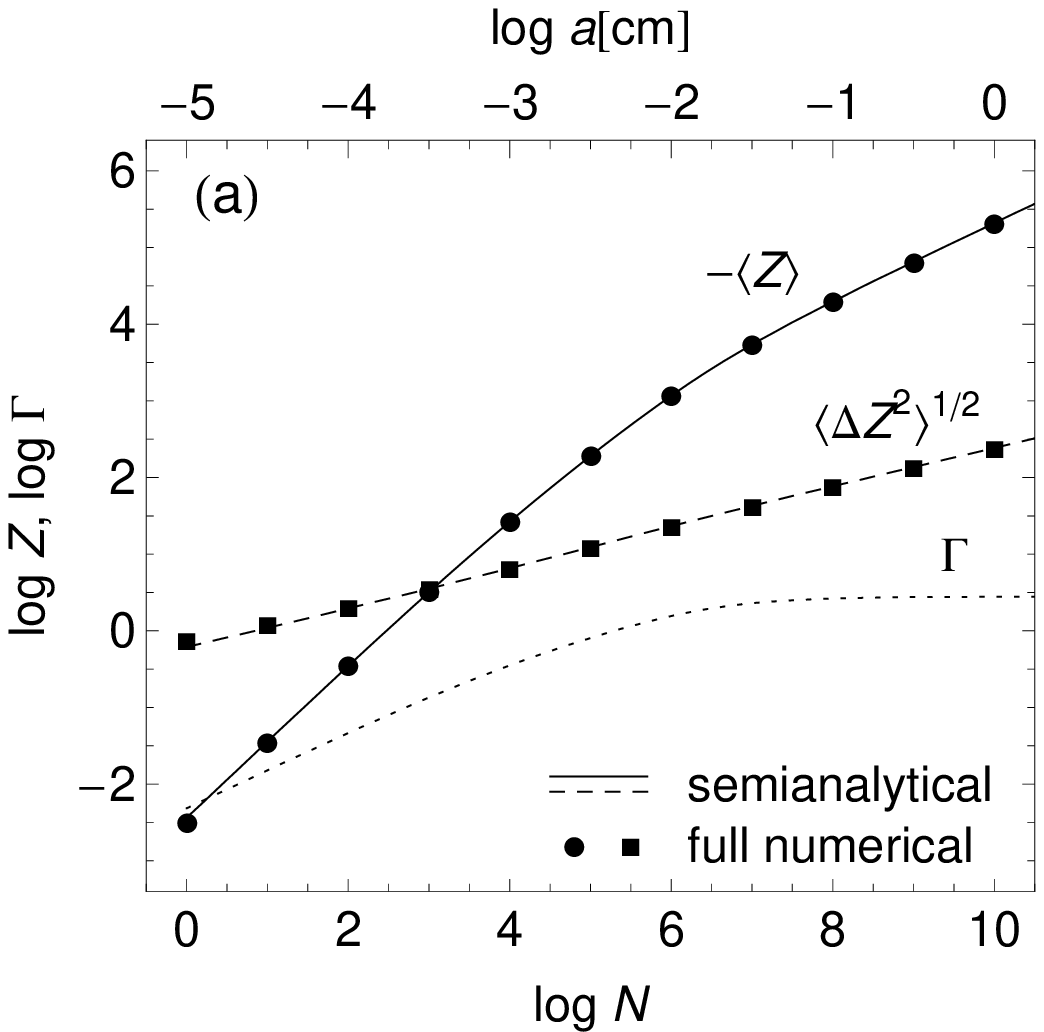}{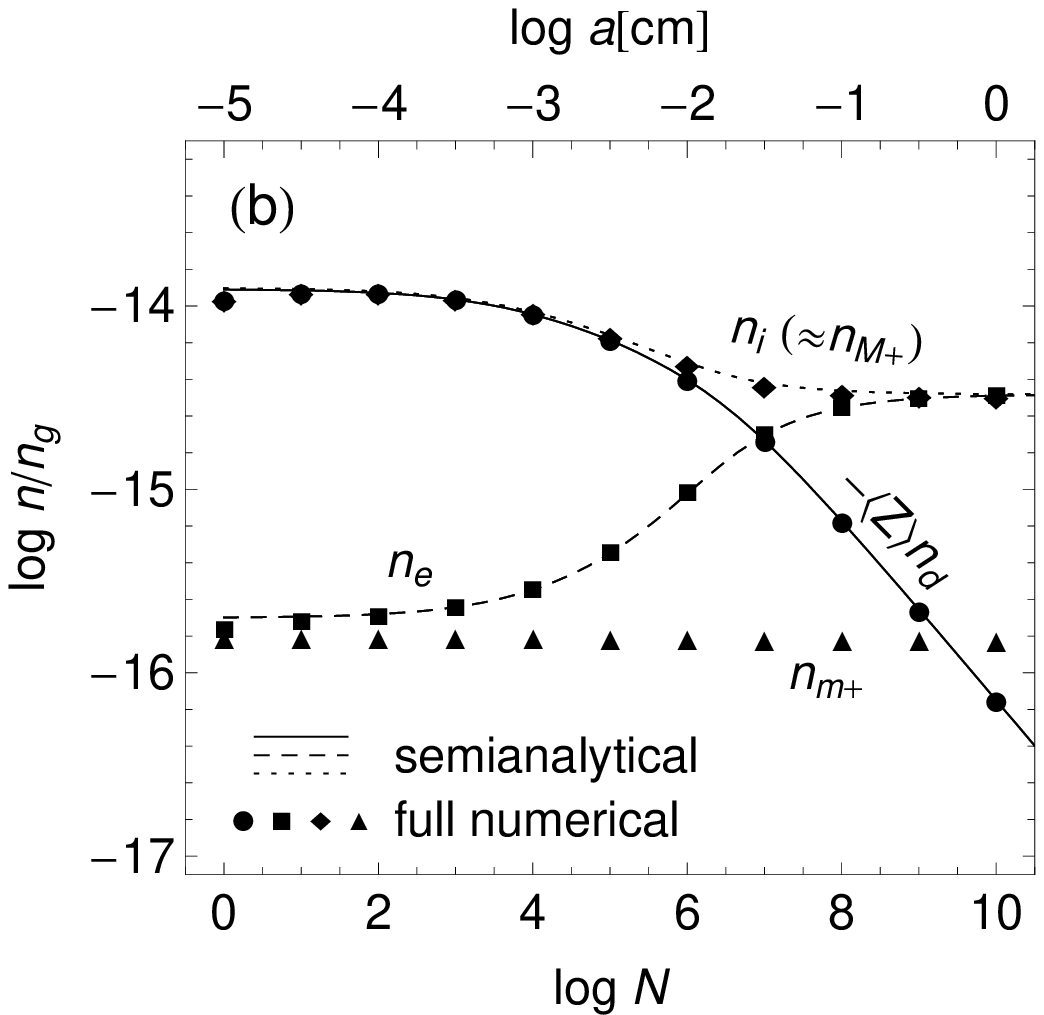}
\caption{({\it a}) The evolution of the dust charge distribution for the ``BCCA'' (i.e., quasi-monodisperse, $D\approx2$) dust growth. 
The solid, dashed, and dotted curves represent the average $\bracket{Z}$,
dispersion $\bracket{\Delta Z^2}^{1/2}$, and master parameter $\Gamma$
of the charge distribution calculated using the semianalytical method 
(eqs.[\ref{eq:Zavr}], [\ref{eq:Zvar}], and [\ref{eq:neut_G}]).
The filled circles and squares denote $\bracket{Z}$ and  $\bracket{\Delta Z^2}^{1/2}$
obtained from full-numerical calculations using the Umebayashi \& Nakano model (see \S3.1.3).
({\it b}) The evolution of the gas ionization state and the dust charge state for the BCCA growth.
$n_\si$ and $n_\se$ are the number densities of ions and electrons in the gas phase, and
$\bracket{Z}n_\sd$ is the net dust charge density.
$n_{\rm M^+}$ and $n_{\rm m^+}$ are the number densities of metal ions and molecular ions.
}
\label{fig:Zx}
\end{figure*}
%%%%%%%%%%%%
Here we show the result for the case $\alpha_{\rm turb} = 0$ (i.e., laminar disk) and $a_0=0.1\micron$
as a fiducial example.
The results for different values of $a_0$ and $\alpha_{\rm turb}$ 
are examined in \S3.2.2 and \S3.2.3, respectively.

Figure \ref{fig:Zx} shows the evolution of dust charge state and gas ionization state
at $r = 5\,\AU$, $z=H$.
Here each evolutionary stage is labeled by the number of constituent monomers in an aggregate, $N$.
The evolution of the mean $\bracket{Z}$ and dispersion $\bracket{\Delta Z^2}$ of
the dust charge distribution as well as the master parameter $\Gamma$ is shown 
in figure \ref{fig:Zx}{\it a}.
We find that $\Gamma$ increases with $N$ and reaches to 
the maximum value $\Gamma_{\rm max}=2.8$ at $N\approx 10^7$.
This means that the gas-dust mixture is an ion-dust plasma ($\Theta \ll \Gamma_{\rm max}$) 
at the initial stage, 
and evolves into an ion-electron plasma ($\Theta \gg \Gamma_{\rm max}$) as the dust grows.
This is expected from the analysis in the last section:
in the BCCA growth, $\Theta \propto 1/(\sigma a n_\sd^2)$ is
proportional to $N^{1/2}$, and thus increases with the growth.
We have confirmed that the above value of $\Gamma_{\rm max}$ 
is consistent with the solution to equation \eqref{eq:ion-elec}.

It is useful to introduce the critical monomer number $N\equiv N_{\rm max}$ 
at which the transition from the ion-dust plasma regime to the ion-electron plasma regime occurs.
As explained in \S2.3, this value can be 
estimated by setting $\Theta \approx \Gamma_{\rm max}$, or
\beq
\sigma a n_\sd^2 \approx 
\frac{\zeta n_\sg \lambda}{u_\si \Gamma_{\rm max}}.
\eeq
Substituting $n_\sd \approx n_{\sd0}/N$, $\sigma \approx \sigma_0N$, and $a\approx a_0N^{1/2}$ 
into this equation,
we obtain the critical size for the transition
\beqn
N_{\rm max} &\approx& \left( \dfrac{\sigma_0 a_0 u_\si n_{\sd0}^2\Gamma_{\rm max}}{\zeta n_\sg e^2}
\right)^{1/2} \nonumber \\
&\approx& 10^6 f_\Sigma^2\pfrac{f_{\rm dg}}{0.014}^{4}
\pfrac{r}{5\,\AU}^{-6}\pfrac{T}{130\,{\rm K}}^{2}\pfrac{\zeta}{10^{-17}{\rm /s}}^{-2}
 \nonumber \\
&&\times \pfrac{a_0}{0.1\micron}^{-6}\pfrac{\rho_0}{1.4{\rm g/cm^3}}^{-4},
\label{eq:Nmax}
\eeqn
or equivalently,
\beqn
a_{\rm max} &\approx& a_0 N_{\rm max}^{1/2} \nonumber \\
&\approx& 10^{-2} f_\Sigma\pfrac{f_{\rm dg}}{0.014}^2
\pfrac{r}{5\,\AU}^{-3}\pfrac{T}{130\,{\rm K}}\pfrac{\zeta}{10^{-17}{\rm /s}}^{-1/2}
 \nonumber \\
&&\times \pfrac{a_0}{0.1\micron}^{-2}\pfrac{\rho_0}{1.4{\rm g/cm^3}}^{-2} {\rm cm},
\label{eq:amax}
\eeqn
where we have explicitly expressed the dependence on $f_\Sigma$.

The mean (negative) charge $-\bracket{Z}$ is proportional to $N$ in the ion-dust regime
$(N \ll N_{\rm max})$, and is proportional to $N^{1/2}$ in the ion-electron regime
$(N \gg N_{\rm max})$. 
This is easily understood  if one recalls that $-\bracket{Z} \propto \Gamma a \propto \Gamma N^{1/2}$.
For $N \ll N_{\rm max}$, $\Gamma \approx \Theta$ is proportional to $N^{1/2}$, 
so $-\bracket{Z} \propto N$. For $N \gg N_{\rm max}$,  $\Gamma$ approaches a constant, and thus $-\bracket{Z} \propto N^{1/2}$.
On the other hand, the dispersion $\bracket{\Delta Z^2}^{1/2}$ is found to be nearly 
proportional to $N^{1/4}$, which is because $\bracket{\Delta Z^2}^{1/2} \sim (a/\lambda)^{1/2}$ 
and $a \propto N^{1/2}$.
It is important to notice here that the relative width $|\bracket{\Delta Z^2}^{1/2}/\bracket{Z}|$ 
of the charge distribution becomes sharper and sharper as the dust grows.

The transition of the plasma state is better illustrated by figure \ref{fig:Zx}{\it b}.
This figure shows the number densities of ions and electrons, $n_\si$ and $n_\se$,
as well as the net dust charge density $\bracket{Z}n_{\sd}$. 
For $N \ll N_{\rm max}$, dust is the dominant carrier of negative charges,
as expected for the ion-dust plasma state.
The free electron density $n_\se$ is smaller than that of ions by a factor of 
$(1/s_\se)(m_\se/m_\si)^{1/2} \sim 10^{-2}$ (see eq.~[\ref{eq:xe/xi}]).
As the aggregates grow and $N$ reaches the critical number $N_{\rm max}$, 
the dust charge density $-\bracket{Z}n_\sd$ begins to decrease and $n_\se$ begins to increase.
Finally, at $N\approx 10^{7}$, free electrons become the dominant negative charge carrier,
and the ion-electron plasma state $(n_\si \approx n_\se)$ is established.  

Interestingly, the abundances of charged species, $n_\si$ and $n_\se$, are nearly constant
for both limits of ion-dust and ion-electron plasma regimes.
This result is in contrast to that of previous studies based on the classical, compact ($D=3$) growth model 
(e.g., \citealt*{Sano+00,Wardle07}) in which $n_{\si}$ and $n_{\se}$ increase as the dust grows.
This difference is attributed to the fact 
that the net projected area $\sigma n_{\sd}$ of dust aggregates
is kept nearly constant for $D\la 2$,
while it decreases for $D=3$.
Using the constancy $\sigma n_{\sd} = \sigma_0 n_{\sd0}$, 
equations \eqref{eq:xi_G} and \eqref{eq:xe_G} can be rewritten as 
$n_\si = n_{\si0}/({1+\Gamma})$ and $n_\se = n_{\si0}(s_\se u_\se/u_\si)\exp\Gamma$,
where \beqn
n_{\si0} &=& \frac{\zeta n_\sg}{u_\si\sigma_0n_{\sd0}} \nonumber \\
&\approx& 10^{-14}\pfrac{f_{\sd\sg}}{0.014}^{-1}\pfrac{r}{5\AU}^3
\pfrac{\zeta}{10^{-17}{\rm/s}} \nonumber \\
&&\times\pfrac{a_0}{0.1\micron}^{-3}\pfrac{\rho_0}{1.4{\rm g/cm^3}}^{-1}n_\sg
\eeqn
is the abundance of ions for $N \ll N _{\rm max}$.
For both plasma limits, the factors $1+\Gamma$ and $\exp\Gamma$ are approximately constant, 
so both $n_\si$ and $n_\se$ approach constant values.
Physically speaking, the constancy of $\sigma n_\sd$ for $D\la 2$ means 
that all the monomers in a fractal aggregate with $D\la 2$ are exposed to outer space, and are thus 
capable to capture free electrons and ions. 
 
Figure \ref{fig:Zx} also shows the result of full numerical calculations 
using the simplified ion-reaction scheme.
We find an excellent agreement between the semianalytical and full numerical calculations.
It is clear that our semianalytical approach is not only efficient but also accurate. 
The most abundant ions are metal ions for all stages of dust evolution.
Molecular ions, the second most abundant ones, are an order of magnitude fewer than metal ions.   

%%%%%%%%%%%%%%
\begin{figure}
\plotone{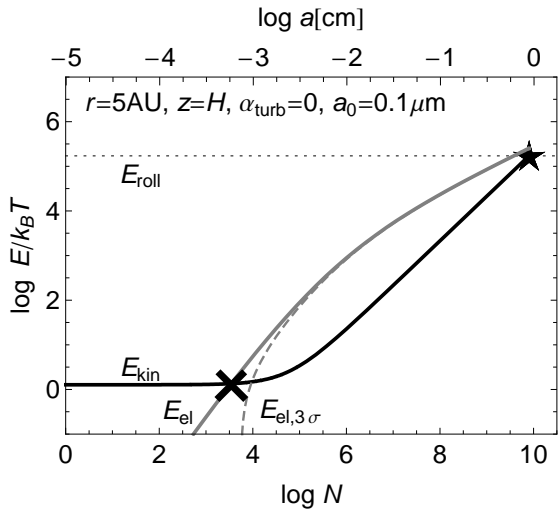}
\caption{
The relative kinetic energy $E_{\rm kin}$ (eq.~[\ref{eq:Ekin}]; solid black curve) versus
the electrostatic energy $E_{\rm el}$ (eq.~[\eqref{eq:Eel}]; solid gray curve)
for the BCCA growth as a function of the monomer number $N$.
The disk position is $(r,z)=(5{\rm AU},H)$, and the turbulence parameter and the monomer size are 
set to $\alpha_{\rm turb}=0$ (i.e., laminar) and $a_0=0.1\micron$. 
The collision cross section for an aggregate pair with $E_{\rm el}$ and $E_{\rm kin}$ 
 vanishes when $E_{\rm el}>E_{\rm kin}$ (see eq.~[\ref{eq:growthcond}]),
meaning that the BCCA growth ``freezes out'' at the size indicated by the cross ($\times$) symbol.
The dashed gray curve shows the ``three-sigma'' electrostatic energy $E_{\rm el,3\sigma}$ (eq.~[\ref{eq:Eel_3sigma}]), representing the effect of charge fluctuation.
The star ($\star$) symbol indicates the size at which $E_{\rm kin}$ reaches the critical rolling-friction energy $E_{\rm roll}$ (eq.~[\ref{eq:Eroll}]; dotted black curve).
Above this critical energy, collisional compression of aggregates becomes effective.
}
\label{fig:EN}
\end{figure}
%%%%%%%%%%%
Now we examine the growth condition.
Figure \ref{fig:EN} shows the kinetic energy $E_{\rm kin}$
and electrostatic energy $E_{\rm el}$ for colliding aggregates at each evolutionary stages.
For $N \la 10^5 \,(a\la 30\micron)$, the thermal (Brownian) motion dominates the relative velocity of 
the aggregates, so $E_{\rm kin}$ is kept constant $\approx \kB T$.
Fore $N \ga 10^5$,  the vertical sedimentation dominates the aggregate motion and
 $E_{\rm kin}$ increases with $N$. 
On the other hand, $E_{\rm el}$ always grows with $N$, 
in proportional to $N^{3/2}$ for $N\la N_{\rm max}$ and to $ N^{1/2}$ for $N \ga N_{\rm max}$. 
This is explained from the fact that $|\bracket{Z}|\propto N$ for the ion-dust plasma regime
($N\la N_{\rm max}$) and $|\bracket{Z}|\propto a \propto N^{1/2}$ for the ion-electron plasma regime
($N\ga N_{\rm max}$).
As a result, the growth condition \eqref{eq:growthcond} breaks down at $N\approx 10^{3.5}\,(a\approx 6\micron)$.
This means that the collision between aggregates with average charge
 $\bracket{Z}$ becomes {\it impossible} at this stage.
Note that the repulsion energy $E_{\rm el}$ reaches $10$ times the thermal energy $\sim \kB T$
as early as $N \approx 10^4$, and finally goes up to $10^5 \kB T$ at the onset of collisional 
compression. 
It is evident that the thermal fluctuation of the kinetic energy 
cannot help the aggregates to grow beyond $N \gg 10^4$.

One may expect that the fluctuation of aggregate charge could help the growth.
To see this effect, we overplot in figure \ref{fig:EN} the ``three-sigma'' electrostatic energy $E_{\rm el,3\sigma}$ defined in equation \eqref{eq:Eel_3sigma}.
We see that $E_{\rm el,3\sigma}$ quickly converges to $E_{\rm el}$ and finally exceeds $E_{\rm kin}$ 
at $N \approx 10^4$.
This is expected from figure~\ref{fig:Zx}{\it a}: 
the relative width $|\bracket{Z}/\bracket{\Delta Z^2}^{1/2}|$ of the the charge distribution becomes narrower and narrower as the dust grows.
Therefore, the result that the aggregates cannot grow beyond $N\gg 10^4$ is preserved even if 
the charge fluctuation is taken into account.\footnote{
It is also found that $E_{\rm el,3\sigma}$ is negative for smaller sizes, $N \la 10^{3.7}$. 
This means that colliding aggregates can possess opposite charges with a probability larger than ``three-sigma''.
This is because the average negative charge $-\bracket{Z}$ in this stage is smaller than $3\bracket{\Delta Z^2}^{1/2}$, as seen in figure \ref{fig:Zx}{\it a}.
However, the attraction energy $-E_{\rm el,3\sigma}$ is insignificant: at most half of the kinetic energy $E_{\rm kin}$.}
All the above facts suggest that the BCCA growth of dust aggregates at this disk position ``freezes out'' 
at size $N \sim 10^4 (a \sim 10\micron)$.

It is interesting to compare this critical size for the freeze-out with that for the first compression,
i.e., the size at which the growth mode changes from the BCCA to the growth involving collisional compression.
We overplot the critical rolling energy $E_{\rm roll}$ (eq.~[\ref{eq:Eroll}]) in figure \ref{fig:EN}.
Comparing this critical energy with $E_{\rm kin}$, we find that the collisional compression begins 
at size $N\approx 10^{10} (a \approx 1{\rm cm})$, i.e., many orders of magnitude smaller than
the above critical freeze-out size.
This illustrates how fast the charging of aggregates begins to affect their collisional growth.

%%%%%%%%%%%%%
\begin{figure*}
\plottwo{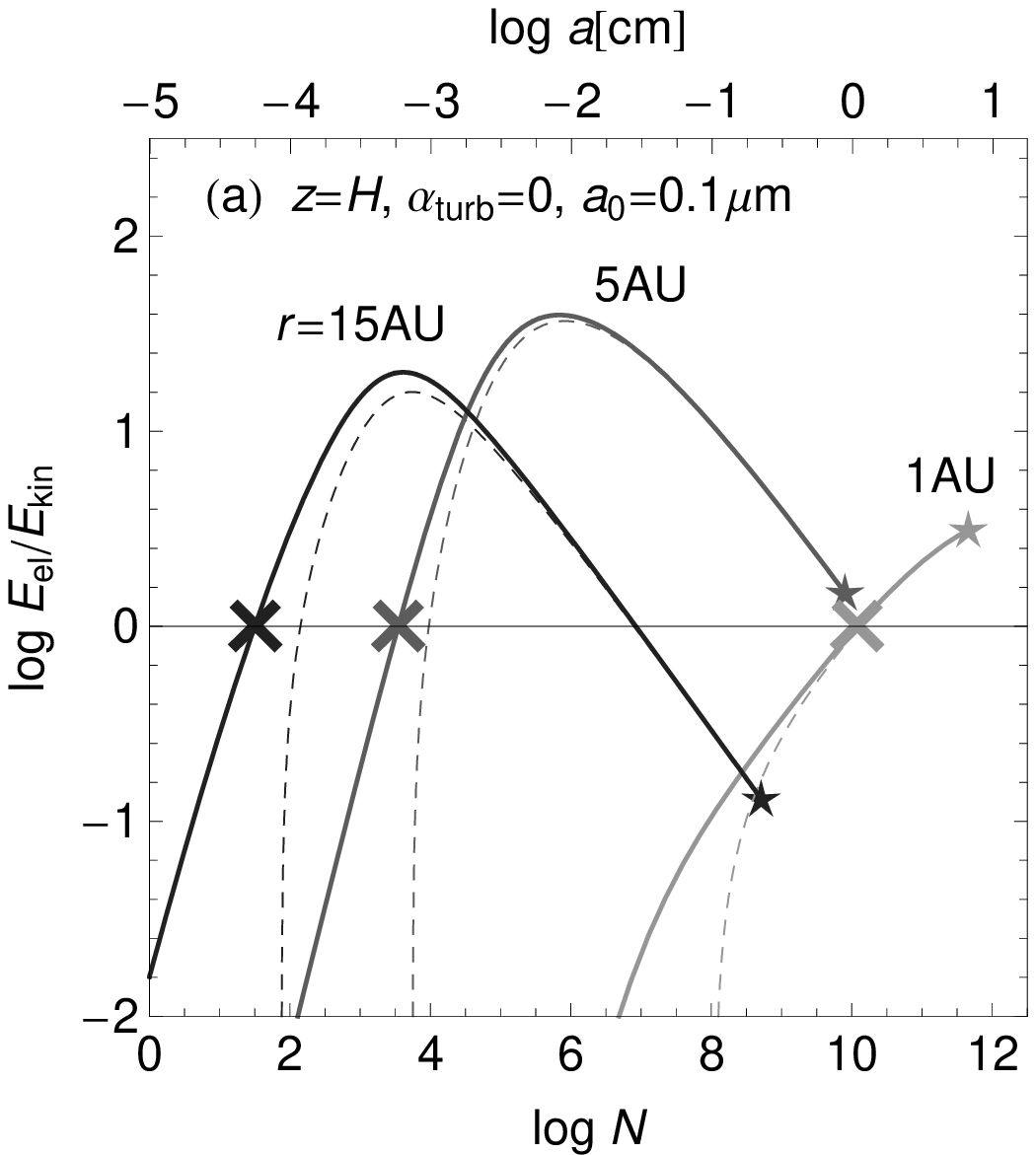}{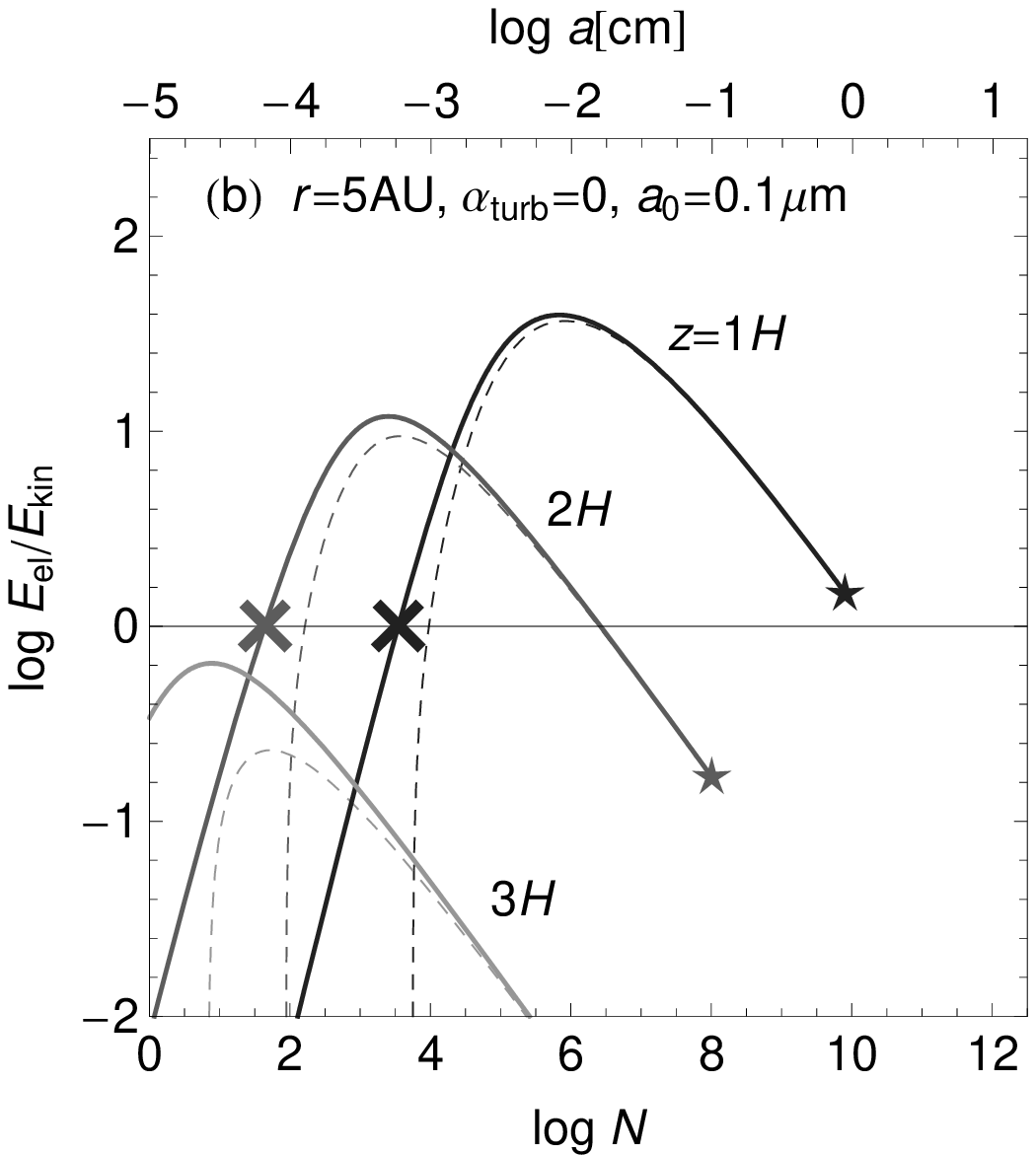}
\caption{
The ratios $E_{\rm el}/E_{\rm kin}$ between the electrostatic and kinetic energies ({\it solid curves})
at various disk radii $r$ ({\it a}) and altitudes $z$ ({\it b}) as functions of $N$. 
The $\alpha$-parameter and the monomer radius 
are set to $\alpha_{\rm turb}=0$ and $a_0=0.1\micron$.
The cross ($\times$) symbols indicate the size at which the growth condition \eqref{eq:growthcond}
breaks down for aggregates with mean charge $\bracket{Z}$.
The dashed curves show the ratio $E_{\rm el,3\sigma}/E_{\rm kin}$
for the ``three-sigma'' energy, which represents the effect of charge dispersion. 
The star ($\star$) symbols indicate the sizes at which $E_{\rm kin}$ reaches the critical rolling-friction energy $E_{\rm roll}$.
} 
\label{fig:Erz}
\end{figure*}
%%%%%%%%%%%%
Figure \ref{fig:Erz} compares the energy ratio $E_{\rm el}/E_{\rm kin}$
for different disk positions. 
We find that the barrier against the BCCA growth appears  
irrespectively of the disk radius $r$.
By contrast, the growth barrier vanishes at $z \ga 3H$ because
the sedimentation velocity at the high altitudes is large enough 
for aggregates to overcome the barrier.

To summarize, the freeze-out of the BCCA growth is very likely to occur in this fiducial model,
except at high altitudes over the midplane.

As seen in figure \ref{fig:EN}, the energy ratio $E_{\rm el}/E_{\rm kin}$ 
takes its maximum at $N\approx N_{\rm max}$.
With this fact, we can roughly estimate the maximum value $(E_{\rm el}/E_{\rm kin})_{\rm max}$
of the energy ratio at $z\approx H$ as follows.
We assume that the relative motion of aggregates by $N\approx N_{\rm max}$ is dominated by vertical sedimentation, as is for $r=5\AU$.
Then, substituting $\Delta u \approx \Delta u_{\rm sed}$ into equation \eqref{eq:Ekin}
and using equations \eqref{eq:Eel} and \eqref{eq:Nmax}, 
we obtain
\beqn
\pfrac{E_{\rm el}}{E_{\rm kin}}_{\rm max}
&\approx& 30 f_\Sigma \pfrac{f_{\rm dg}}{0.014}^{-2}\pfrac{\zeta}{10^{-17}{\rm /s}}
 \nonumber \\
&&\times \pfrac{a_0}{0.1\micron}^{-1}\pfrac{\rho_0}{1.4{\rm g/cm^3}}^{-1}
\label{eq:Eratio}
\eeqn
at $z\approx H$. Notably, $(E_{\rm el}/E_{\rm kin})_{\rm max}$ is explicitly independent of both $r$ and $T$.
This means that the ``height'' of the growth barrier is insensitive to the temperature profile.
Equation \eqref{eq:Eratio} does not hold beyond $r \approx 5 \,\AU$,
since the maximum energy ratio appears in the Brownian motion regime.
This equation is nevertheless useful because it allows us a rough estimation 
on how the growth barrier depend on the model parameters.
For example, for fixed $\zeta$, $a_0$, and $\rho_0$,
 the the growth barrier is more serious if the disk is more massive ($f_\Sigma>1$) 
or more depleted of dust ($f_{\sd\sg}< 10^{-2}$). 
%%%%%%%%%%%%%%%%%%%%%%%%%%%%%%%%%
\subsubsection{Effect of monomer size}
%%%%%%%%%%%%%%%%%%%%%%%%%%%%%%%%%
%%%%%%%%%%%%%
\begin{figure}
\plotone{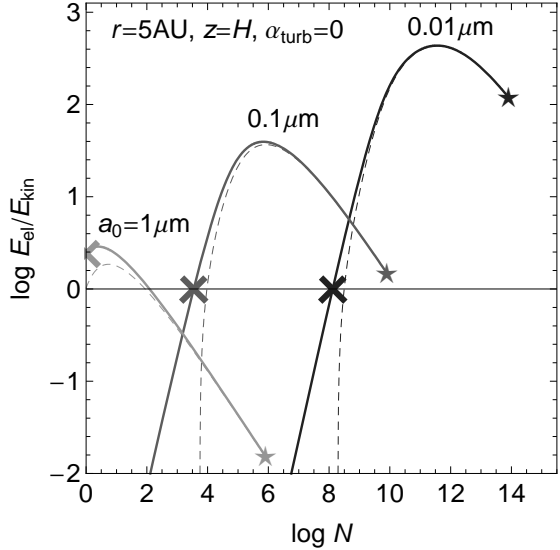}
\caption{
The energy ratios $E_{\rm el}/E_{\rm kin}$ ({\it solid curves}) at $r=5{\rm AU}$ and $z = H$  for different monomer sizes $a_0$.
The cross $(\times)$ symbols indicate the ``freeze-out'' sizes at which the growth condition \eqref{eq:growthcond} breaks down.
The dashed curves show $E_{\rm el,3\sigma}/E_{\rm kin}$.
The star ($\star$) symbols indicate the sizes at which $E_{\rm kin}$ reaches the critical rolling-friction energy $E_{\rm roll}$.
} 
\label{fig:Er1}
\end{figure}
%%%%%%%%%%%%
The actual size of dust monomers in protoplanetary disks is unknown.
Infrared observations of interstellar medium suggest the size distribution of 
interstellar grains ranges from $\approx 0.005\micron$ to $\approx 0.25\micron$
(MRN distribution; {\citealt*{MRN77}).
If interstellar grains are not aggregates but monomers,
the typical monomer size in protoplanetary disks will fall within 
the range $0.001\micron \la a \la 1\micron$.
Figure \ref{fig:Er1} compares the energy ratio $E_{\rm el}/E_{\rm kin}$ 
for different monomer sizes $a_0 = 0.01,\;0.1$, and $1\micron$.
We find that $N_{\rm max} \propto a_0^{-6}$ ($a_{\rm max} \propto a_0^{-2}$) 
and $(E_{\rm el}/E_{\rm kin})_{\rm max} \propto a_0^{-1}$, 
as expected from equations \eqref{eq:Nmax} and \eqref{eq:Eratio}.
This delay (measured in the ``degree of growth'' $N$) is attributed to 
the larger cross section of BCCA clusters composed of smaller monomers.
Such a larger cross section causes a quick depletion of free electrons in the gas phase,
resulting in the delay of transition from the ion-dust plasma regime to the ion-electron regime.
In addition, the larger cross section produces a stronger coupling to the gas, 
and in turn a slower increase of the kinetic energy.  
We find that the growth condition \eqref{eq:growthcond} breaks down much before the onset of collisional compression for all $a_0 \la 1\micron$.
Therefore, the ``freeze-out'' of the BCCA growth is not prevented
even if the monomer size in protoplanetary disks is assumed to the maximum value inferred by the MRN distribution.

%%%%%%%%%%%%%%%%%%%%%%%%%%%%%%%%%
\subsubsection{Effect of turbulence}
%%%%%%%%%%%%%%%%%%%%%%%%%%%%%%%%
%%%%%%%%%%%%%
\begin{figure}
\plotone{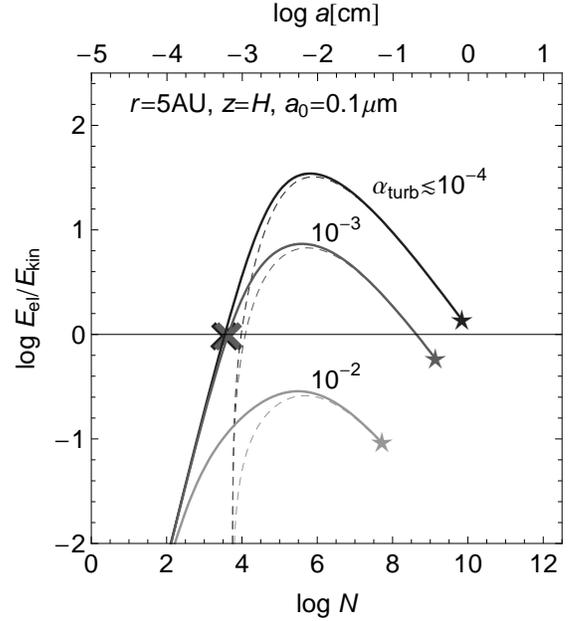}
\caption{
The energy ratios $E_{\rm el}/E_{\rm kin}$ ({\it solid curves}) at $r=5{\rm AU}$ and $z = H$ for different turbulence strengths $\alpha_{\rm turb}$.
The cross $(\times)$ symbols indicate the ``freeze-out'' sizes at which the growth condition \eqref{eq:growthcond} breaks down.
The dashed curves show $E_{\rm el,3\sigma}/E_{\rm kin}$.
The star ($\star$) symbols indicate the sizes at which $E_{\rm kin}$ reaches the critical rolling-friction energy $E_{\rm roll}$.
} 
\label{fig:Er2}
\end{figure}
%%%%%%%%%%%%

Here we examine how strong turbulence is needed to remove the growth barrier.
Figure \ref{fig:Er2} shows the evolution of energy ratio $E_{\rm el}/E_{\rm kin}$
for three turbulent cases $\alpha_{\rm turb}=10^{-4},10^{-3},$ and $10^{-2}$.
We see that turbulence of $\alpha_{\rm turb} \la 10^{-4}$ does not affect the energy ratio
for any size $N$. 
This is not surprising because both $\Delta u_{\rm turb}$ and $\Delta u_{\rm sed}$
scale with $\Delta t_{\rm stop}$.
The relative velocity $\Delta u_{\rm turb}$ induced by turbulence is estimated as
$\Delta u_{\rm turb} \approx \Delta t_{\rm stop}{\rm Re}^{1/4}u_{\rm large}/t_{\rm large}
\sim \Delta t_{\rm stop}{\rm Re}^{1/4}\alpha_{\rm turb}^{1/2}\Omega_{\rm K}^2H$.
At $z \sim H$, the Reynolds number ${\rm Re}$ is of order 
 $\sim \alpha_{\rm turb}\Sigma_\sg\sigma_{\rm mol}/m_\sg$,
so $\Delta u_{\rm turb}$ is written as
\beq
\Delta u_{\rm turb} 
\sim \pfrac{\Sigma_\sg\sigma_{\rm mol}}{m_\sg}^{1/4}\alpha_{\rm turb}^{3/4}
\Delta t_{\rm stop}\Omega_{\rm K}^2H
\eeq
On the other hand, the relative velocity by differential sedimentation is 
$\Delta u_{\rm sed} \sim \Delta t_{\rm stop}\Omega_{\rm K}^2H$, so we find
\beq
\frac{\Delta u_{\rm turb}}{\Delta u_{\rm sed}} 
\sim \pfrac{\Sigma_\sg\sigma_{\rm mol}}{m_\sg}^{1/4}\alpha_{\rm turb}^{3/4}
\sim \pfrac{\alpha_{\rm turb}}{10^{-4}}^{3/4}\pfrac{r}{5\,\AU}^{-3/8}.
\eeq 
Therefore, the effect of turbulence on the aggregate collision
is negligible for all $N$ as long as $\alpha_{\rm turb} \la 10^{-4}$.

For $\alpha_{\rm turb}\ga 10^{-2}$, we find that 
the growth barrier is entirely removed. 
This suggests that relatively strong $(\alpha_{\rm turb}\ga 10^{-2})$ turbulence 
is a key ingredient for early stages of dust coagulation.
In \S4.3, we discuss this topic in more detail.

%%%%%%%%%%%%%%%%%%%%%%%%%%%%%%%%%%%%%%%%
\section{Discussion \label{sec:discussion}}
%%%%%%%%%%%%%%%%%%%%%%%%%%%%%%%%%%%%%%%%
%%%%%%%%%%%%%%%%%%%%%%%%%%%%%%%%%%%%%%%%%%%%%%%%%%%%%%%%%%%%%%%%%%%%%%%%%%%%%
\subsection{Validity of the charge equilibrium}
%%%%%%%%%%%%%%%%%%%%%%%%%%%%%%%%%%%%%%%%%%%%%%%%%%%%%%%%%%%%%%%%%%%%%%%%%%%%%
In this study, we have assumed that the charge reactions are
much faster than the dust-dust collisions.
Now we show that this assumption is actually valid for evolutionary stages which we are interested in.

The typical time scale for the system to relax to an ionization-recombination equilibrium
can be measured by the average time $t_{\rm coll,d}$ needed for an aggregate to collide with an ion,  
\beq
t_{\rm coll,i}^{-1} \approx u_\si \sigma_{\rm di} n_\si \approx \frac{\zeta n_\sg}{n_\sd},
\eeq
where we have used that $\zeta n_\sg \approx u_\si\sigma_{\sd\si} n_\si n_\sd$ since
 the gas-phase recombination is negligible in the presence of dust (see \S3.1.3).
On the other hand, the mean collision time $t_{\rm coll,d}$ between aggregates is written as 
\beq
t_{\rm coll,d}^{-1} \approx \sigma_{\rm dd} n_\sd\Delta u  \sim  \sigma n_\sd \Delta u.
\eeq
Therefore, the ratio of these time scales is estimated as
\beq
\frac{t_{\rm coll, d}}{t_{\rm coll, i}} 
\sim \frac{\zeta n_\sg}{\sigma n_\sd^2 \Delta u} 
\sim f_{\sd\sg}^{-2}\pfrac{m}{m_\sg}^2\frac{c_s}{\Delta u}\frac{m_\sg}{\sigma\Sigma_\sg}\frac{\zeta}{\Omega_{\rm K}}.
\eeq
If the dust velocity is dominated by the Brownian motion, as is for small aggregates,
$\Delta u \sim c_s\sqrt{m_\sg/m}$ and
\beqn
\frac{t_{\rm coll, d}}{t_{\rm coll, i}}\biggr|_{\rm Brown}  
&\sim& f_{\sd\sg}^{-2}\pfrac{m}{m_\sg}^{5/2}\frac{m_\sg}{\sigma \Sigma_\sg}
\frac{\zeta}{\Omega_{\rm K}} \nonumber \\
&\sim& N^{3/2}\pfrac{f_{\sd\sg}}{10^{-2}}^{-2}
 \pfrac{\zeta}{10^{-17}{\rm /s}}\pfrac{r}{2\,\AU}^{3},
\eeqn
where we have used $a_0 \sim 0.1\micron$ and $\rho_0 \sim 1{\rm g/cm^3}$.
We find that dust coagulation can be safely neglected
if $r \ga 2\AU$ or $N \gg 1$.
This is true even if the motion of dust aggregates is dominated 
by vertical sedimentation, since $\Delta u_{\rm sed} \sim c_s(m/\Sigma_\sg\sigma)$, and thus
\beqn
\frac{t_{\rm coll, d}}{t_{\rm coll, i}}\biggr|_{\rm sed} 
&\sim& f_{\sd\sg}^{-2}\frac{m}{m_\sg}\frac{\zeta}{\Omega_{\rm K}} \nonumber \\
&\sim& 10^3 N \pfrac{f_{\sd\sg}}{10^{-2}}^{-2}\pfrac{\zeta}{10^{-17}{\rm /s}}\pfrac{r}{2\,\AU}^{3/2}.
\eeqn
Thus, it is concluded that the growth of dust aggregates 
can be neglected if $r \ga 2\AU$ or $N \gg 1$.

%%%%%%%%%%%%%%%%%%%%%%%%%%%%%%%%%%%%%%%%%
\subsection{Internal electrostatic force}
%%%%%%%%%%%%%%%%%%%%%%%%%%%%%%%%%%%%%%%%%
In the last section, we have implicitly assumed that charged aggregates can stick to each other
as long as the collision condition \eqref{eq:growthcond} is satisfied.
One might wonder if the collided aggregates are pulled off from each other by the electrostatic repulsion.
In fact, this repulsion is much weaker than the attraction (due to van der Waals force) between two monomers in contact. 
The electrostatic repulsion force $F_{\rm el,int}$ acting between two collided aggregates is estimated as  
\beq
F_{\rm el,int} \sim \frac{(Ze)^2}{a^2} \sim \frac{\Gamma^2 e^2}{\lambda^2} 
\la 10^{-8} \pfrac{T}{130\,{\rm K}}^2\;{\rm dyn},
\eeq
where we have used that $\Gamma<\Gamma_{\rm max} \sim 3$.
Note that maximum value of $F_{\rm el,int}$ is independent of the aggregate size $a$.
On the other hand, the critical force needed to separate two monomers in contact is 
\citep{JKR71,DT97}
\beq
F_{\rm crit} = 3\pi\gamma \frac{a_0}{2} \sim 10^{-3}\pfrac{\gamma}{10^2{\rm g/cm^2}}\pfrac{a_0}{0.1\micron}\;{\rm dyn},
\eeq
where $\gamma$ is the surface adhesive energy mentioned in the last section.
Thus, the electrostatic force inside an aggregate is negligibly weak compared to the contact force between two constituent monomers.

%%%%%%%%%%%%%%%%%%%%%%
\subsection{Dust growth in strong turbulence}
%%%%%%%%%%%%%%%%%%%%%%
As seen in \S3.2.3, the electrostatic barrier against the fractal growth will be removed 
if considerably strong $(\alpha_{\rm turb} \ga 10^{-2})$ turbulence is present.
We here discuss whether such turbulence is likely to occur in protoplanetary disks, 
and what would happen after the dust overcome the electrostatic growth barrier.

The most robust mechanism for disk turbulence is MRI
  \footnote{It is unknown whether any mechanism other than MRI can drive and sustain turbulence 
  with $\alpha_{\rm turb} \ga 10^{-2}$ in the early stage of dust evolution.
  For example, convective instability may operate in this stage \citep{LP80}, but it is unlikely to
  sustain such strong turbulence \citep{SB96}.}.
MRI-driven turbulence will achieve $\alpha_{\rm turb} \sim 10^{-2}$ 
in its saturated state (e.g. \citealt*{SIM98}).
Therefore, fractal aggregates will be able to overcome the electric barrier in MRI-active regions.
\citet{Sano+00} calculated the active region using the MMSN model and found that MRI will be active only 
at outer ($r\ga 20{\rm AU}$) disk radii or high ($|z|\ga 2H$) altitudes if the dust size is $0.1\micron$.
The size of the active region does not vary even if the dust grows since, as seen in \S3.2.1, 
the ionization fraction is kept nearly constant as long as the dust undergoes the fractal growth.
Therefore, the region in which the fractal aggregates can overcome the electrostatic barrier 
is limited to outer disk radii and high altitudes.
 
A more serious problem is that such strong turbulence causes another kind of growth barrier,
i.e., catastrophic fragmentation of colliding aggregates.
In turbulent regions, large aggregates with $t_{\rm stop}\sim 1/\Omega_{\rm K}$
have the maximum collisional velocity of order $u_{\rm large} \sim \sqrt{\alpha_{\rm turb}}c_s$, 
which amounts to more than $100{\rm m/s}$ for $\alpha_{\rm turb} \ga 10^{-2}$.
On the other hand, as shown by recent $N$-body simulations \citep{Wada+08},  
catastrophic fragmentation will take place for relative velocity $\Delta u \ga 30 {\rm m/s}$.
Therefore, it is very likely that strong turbulence destroys the aggregates and consequently 
prevents further dust growth.
This idea is supported by a recent statistical study \citep{BDH08}.

Thus, the combination of electric repulsion and collisional fragmentation might 
strictly limit the dust growth and subsequent planetesimal formation in protoplanetary disks.
It is important to think of a possibility that dust evolution will continue in some way
even if the turbulence is weak and the quasi-monodisperse fractal growth does freeze out.
This is the topic of the next subsection.

%%%%%%%%%%%%%%%%%%%%%%%%%%%%%%%%%%%%%%%%%%%%%%%%%%%%%%%%%%%%%%%%%%%%%%%%%%
\subsection{A possible scenario to overcome the electric growth barrier}
%%%%%%%%%%%%%%%%%%%%%%%%%%%%%%%%%%%%%%%%%%%%%%%%%%%%%%%%%%%%%%%%%%%%%%%%%%
In \S3, we have ignored the size distribution of aggregates.
In fact, there may exist some aggregates considerably larger than average-sized ones.
In the following, we consider whether such large aggregates can continue to grow even if
 the growth of average-sized ones has frozen out.

Let us consider a small population of irregularly large aggregates
(referred to as ``test aggregates'') growing with a large population of standard ($D \sim 2$) 
fractal aggregates (``field aggregates'').
Under this assumption, the kinetic energy of relative motion between test and field aggregates
is written as
\beq
E_{\rm kin,tf} = \frac{1}{2}\frac{m_{\rm t} m_{\rm f}}{m_{\rm t} + m_{\rm f}}
(\Delta u_{\rm tf})^2 \approx \frac{1}{2}m_{\rm f} (\Delta u_{\rm tf})^2 
= \frac{1}{2}\pfrac{\Delta u_{\rm tf}}{\Delta u_{\rm ff}}^2E_{\rm kin,ff},
\eeq
where the subscripts `t' and `f' respectively represent the test and field aggregates, 
and we have used the assumption $m_{\rm t} \gg m_{\rm f}$.
$E_{\rm kin, ff}$ and $\Delta u_{\rm ff}$ are the kinetic energy of relative motion 
and the relative velocity between two field aggregates,
and are thus equivalent to $E_{\rm kin}$ and $\Delta u$ in \S3.
On the other hand, the electrostatic energy between test and field aggregates is
\beq
E_{\rm el,tf} = \frac{a_{\rm t}a_{\rm f}\Gamma^2e^2}{a_{\rm t}+a_{\rm f}} 
\approx \frac{\Gamma^2 a_{\rm f}}{\lambda^2} = \frac{1}{2}E_{\rm el,ff},
\eeq
where $E_{\rm el, ff}$ is equivalent to $E_{\rm el}$ in \S3,
and we have used that $a_{\rm t} \gg a_{\rm f}$.
Thus, the energy ratio $E_{\rm el,tf}/E_{\rm kin,tf}$ is written as
\beq
\frac{E_{\rm el,ft}}{E_{\rm kin,ft}}
 \approx \frac{E_{\rm el,ff}}{E_{\rm kin,ff}}\pfrac{\Delta u_{\rm ff}}{\Delta u_{\rm tf}}^2.
\eeq

Now we assume that the growth of field aggregates has frozen out
due to the electric barrier, i.e., $E_{\rm kin,ff} = E_{\rm el,ff}$.
At this stage, the condition for the collision between test and field aggregates,
$E_{\rm kin,tf}>E_{\rm el,tf}$, reduces to a simple inequality
\beq
\Delta u_{\rm tf} > \Delta u_{\rm ff}.
\label{eq:growthcond_tf}
\eeq

We readily notice that Brownian motion does not satisfy this condition
since $\Delta u_{\rm tf} \approx \sqrt{8 \kB T/\pi m_{\rm f}}$ and 
$\Delta u_{\rm ff} \approx \sqrt{16 \kB T/\pi m_{\rm f}}$.
The remained possibilities are the differential sedimentation and turbulent-driven motion.
In both cases, the relative velocity $\Delta u$ is proportional to $\Delta t_{\rm stop}$, or
$\Delta(m/\sigma)$. 
It is very important to notice here that the condition \eqref{eq:growthcond_tf} is
not safely satisfied as long as the test aggregate is as fluffy as field aggregates, i.e., 
$m_{\rm t}/\sigma_{\rm t}$ is comparable to $m_{\rm f}/\sigma_{\rm f}$.
Hence, the condition \eqref{eq:growthcond_tf} will be safely satisfied
only if the test aggregate is more compact and has larger $m/\sigma$ than the field aggregates.
Moreover, collision with a smaller aggregate generally tends to increase
$m_{\rm t}/\sigma_{\rm t}$, allowing the test aggregate the next collision. 
Therefore, if there exists an aggregate that is large and compact, it will be able to continue growing
by sweeping up smaller ``frozen'' aggregates.

The above consideration suggests that the freeze-out of the quasi-monodisperse fractal growth 
may not mean the termination of dust evolution.
Rather, it may be the beginning of {\it bimodal growth} in which only a small fraction of aggregates
can grow larger and larger while the rest remain frozen.
We plan to examine this possibility in more detail in the future studies.
In any case, we expect that the effect of dust charging should qualitatively modify the current scenario of
dust growth in protoplanetary disks.
  
%%%%%%%%%%%%%%%%%%%%%
\section{Summary}
%%%%%%%%%%%%%%%%%%%%
In this study, we have investigated the electric charging of dust aggregates and its 
effect on collisional dust growth in protoplanetary disks.
We have found that the conditions for ionization-recombination equilibrium
are reduced to a single equation (eq.[\ref{eq:neut_G0}]).  
Just by solving this equation numerically, the dust charge state and gas ionization state 
can be analytically computed for an arbitrary ensemble of aggregates in a self-consistent way.
It is also confirmed that our semianalytical method reproduces the results of a 
previously used, more complicated numerical method (\S3.2.1, fig.3).
This formalism thus provides a fast charge-state solver that will allow  
a coupled simulation of MRI-driven turbulence and dust coagulation.
 
As an application, we have explored the effect of electrostatic charging on
an early stage of dust coagulation in protoplanetary disks.
We considered the quasi-monodisperse fractal growth with the fractal dimension $D\sim2$
as suggested by previous laboratory experiments and $N$-body simulations \citep{Blum04,DBCW07}.
Our findings are summarized as follows:
 
1.
For a wide range of model parameters, the effective cross section for the mutual 
collision of aggregates is quickly suppressed as the fractal growth proceeds 
and finally vanishes at a certain aggregate size (\S\S 3.2.1, 3.2.2).
This is due to the strong electrostatic repulsion between aggregates charging negatively on average,
and happens much before the collisional compression of aggregates becomes effective. 
Both the charge fluctuation and the thermal velocity fluctuation
do not help the aggregates to overcome the growth barrier.
Without strong turbulence, the quasi-monodisperse fractal growth is very likely 
to ``freeze out'' on its way to the subsequent growth stage.

2. 
Strong ($\alpha_{\rm turb} \ga 10^{-2}$) turbulence will help the aggregates to
overcome the above growth barrier (\S3.2.3).
However, such turbulence is likely to occur only in MRI-active regions, i.e., 
at outer disk radii or high altitudes (\S4.3).
Furthermore, it will cause another serious problem---the catastrophic disruption
 of collided aggregates---in later stages.
These facts suggest that the combination of electric repulsion and collisional disruption
may strictly limit the collisional growth of dust aggregates in protoplanetary disks.
 
3. 
The freeze-out of the fractal growth might be followed by bimodal growth 
in which only a small fraction of large aggregates 
can continue growing while a large fraction of small fractal aggregates remains frozen (\S4.4).
This could qualitatively change the current scenario of planetesimal formation in
protoplanetary disks \citep{DBCW07}. 
We will examine this possibility in more detail in forthcoming papers. 

Finally, we point out that the fractal ($D\la 2$) dust growth tends to keep the ionization degree 
of the disk small due to the open nature of aggregates (\S3.2.1, fig.~3{\it b}). 
This means that the magnetorotationally unstable region hardly expands 
until the collisional compression of the aggregates begins to work.
This conclusion is in contrast to that of previous studies (e.g., \citealt*{Sano+00,Wardle07})
which claimed that the ionization degree increase as the aggregates grow.
However, they assumed compact dust growth,
which clearly contradicts recent laboratory experiments and $N$-body simulations.
Thus, the magnetorotational stability of protoplanetary disks
must be reexamined taking into account that the fractal nature of dust aggregates.

%%%%%%%%%%%%%%%%
\acknowledgments
%%%%%%%%%%%%%%%%
The author thanks M. Sakagami, S. Inutsuka, and H. Tanaka for careful reading of the manuscript
and for valuable comments. 

%%%%%%%%%
\appendix
%%%%%%%%%
%%%%%%%%%%%%%%%%%%%%%%%%%%%%%%%%%%%%%%%%%%%%%%%
\section{Derivation of the charge distribution}
%%%%%%%%%%%%%%%%%%%%%%%%%%%%%%%%%%%%%%%%%%%%%%%
Since the velocity of free electrons is much greater than that of ions,
dust aggregates charge up negatively on average. 
Thus, let us assume $Z<0$ and use the expressions of 
$\sigma_{\sd\si}$ and $\sigma_{\sd\se}$ valid for $Z<0$. 
Then, equation \eqref{eq:detailed_xd} is written as
\beq
s_\si u_\si n_\si \left(1-\frac{Z}{\tau}\right)n_\sd(I,Z)
 = s_\se u_\se n_\se \exp\left(\frac{Z+1}{\tau}\right)n_\sd(I,Z+1), 
\label{eq:balance_append}
\eeq
where $\tau = a/\lambda$.

We make the following assumption:
\beq
\frac{1}{\bracket{\Delta Z^2}^{1/2}} \sim \frac{\bracket{\Delta Z^2}^{1/2}}{\tau} \sim \varepsilon,
\label{eq:approx1}
\eeq
where $\bracket{\Delta Z^2}$ is the variance of the charge state distribution, and $\varepsilon \ll 1$.
We also assume that $n_\sd(I,Z)$ varies with a typical scale $\sim \bracket{\Delta Z^2}^{1/2}$. 
Under these assumptions, $n_\sd(I,Z+1)$ can be written as
\beq
n_\sd(I,Z+1) = n_\sd(I,Z) + \frac{\partial n_\sd}{\partial Z}(I,Z) + O(\varepsilon^2).
\eeq
Also, $\exp[(Z+1)/\tau]$ is written as $\exp(Z/\tau) + O(\varepsilon^2)$
since $\tau^{-1} = \bracket{\Delta Z^2}^{-1/2}\cdot\bracket{\Delta Z^2}^{1/2}/\tau \sim \varepsilon^2$.
Substituting them into equation \eqref{eq:balance_append},
we obtain a first-order differential equation for $n_\sd(I,Z)$, 
\beq
\frac{\partial n_\sd}{\partial Z}(I,Z) +W(I,Z)n_\sd(I,Z) \approx 0,
\label{eq:ODE_xd}
\eeq
where
\beq
W(I,Z) \equiv 1- \frac{s_\si u_\si n_\si (1-Z/\tau)}{s_\se u_\se n_\se \exp(Z/\tau)}.
\eeq
Equation \eqref{eq:ODE_xd} is accurate to terms of the first order in $\varepsilon$.

Let us denote the solution of $W(I,Z)=0$ by $Z_0$ and write $Z = Z_0 + \delta Z$.
Also, we make an additional approximation that $|\delta Z|\ll \tau$. 
Expanding $W(I,Z)$ in powers of $|\delta Z|/\tau$ and using $W(I,Z_0)=0$, we have 
\beq
W(I,Z) \approx \frac{2-Z_0/\tau}{1-Z_0/\tau}\frac{\delta Z}{\tau},
\label{eq:W_approx}
\eeq
which is accurate to the first order in $|\delta Z|/\tau$.
Hence, equation \eqref{eq:ODE_xd} is approximated by
\beq
\frac{\partial n_\sd}{\partial Z}(I,Z) 
+ \frac{\delta Z}{\bracket{\Delta Z^2}}n_\sd(I,Z) \approx 0,
\label{eq:ODE_xd_approx}
\eeq
where
\beq
\bracket{\Delta Z^2} \equiv \frac{1-Z_0/\tau}{2-Z_0/\tau}\tau.
\label{eq:Zvar_appendix}
\eeq
It is an easy task to show that the solution to \eqref{eq:ODE_xd_approx} is a Gaussian 
distribution \eqref{eq:Gaussian} with average $\bracket{Z} = Z_0$ and variance $\bracket{\Delta Z^2}$.
Rewriting $Z_0$ and $\tau$ using $Z_0 = -\Gamma \tau$ and $\tau = a/\lambda$, 
we obtain equations \eqref{eq:Zavr} and \eqref{eq:Zvar}.
We note that equation \eqref{eq:Zvar_appendix} has been also obtained by \citet{DS87}, but they did not
show the derivation of this equation in their paper. 

Now we justify the assumption \eqref{eq:approx1}.
The first part of the assumption $\bracket{\Delta Z^2}^{-1/2} \sim \bracket{\Delta Z^2}^{1/2}/\tau$,
which is equivalent to $\bracket{\Delta Z^2} \sim \tau$,
 is always satisfied since $\tau/2<\bracket{\Delta Z^2} < \tau$ (see eq.[\ref{eq:Zvar_appendix}]).
The second part $\bracket{\Delta Z^2}^{1/2}/\tau \sim \varepsilon \ll 1$ is also satisfied 
in typical protoplanetary disks
if dust aggregates are not as small as constituent monomers ($\sim 0.1\micron$), because 
\beq
\frac{\bracket{\Delta Z^2}^{1/2}}{\tau} \sim \tau^{-1/2} 
\sim \left(\frac{a}{0.1\micron}\right)^{-1/2}\left( \frac{T}{130 {\rm K}}\right)^{-1/2}.
\eeq

It is noted that the approximation $|\delta Z| \ll \tau$ used to obtain equation \eqref{eq:W_approx}
is rewritten as $|\delta Z| \ll \tau^{1/2}\bracket{\Delta Z}^{1/2}$. 
Hence, this approximation is good  as long as the region $|\delta Z| \la \bracket{\Delta Z}^{1/2}$ is considered.

%%%%%%%%%%%%%%%%%%%%%%%%%

\end{document}